\documentclass[runningheads]{llncs}
\usepackage[T1]{fontenc}
\usepackage{graphicx}
\usepackage{amsmath,amssymb,amsfonts}
\usepackage{algorithmic}
\usepackage{algorithm}
\usepackage{booktabs}
\usepackage{multirow}
\usepackage{array}
\usepackage{xcolor}
\usepackage{hyperref}
\usepackage{cleveref}
\usepackage{subcaption}
\usepackage{balance}
\usepackage{util}

\newcommand{\ie}{\textit{i.e.}}
\newcommand{\eg}{\textit{e.g.}}

\newcommand{\tow}{t_{\text{ow}}}
\newcommand{\tfw}{t_{\text{fw}}}
\newcommand{\Dp}{D_p}
\newcommand{\yhat}{\hat{y}}
\newcommand{\ftilde}{\tilde{f}}

\newcommand{\LB}{\mathcal{L}_B}
\newcommand{\LW}{\mathcal{L}_W}
\begin{document}
%

% \title{Unveiling Hidden Vulnerabilities in Public Dataset Protection via Forgery Attack}

\title{Forging the Unforgeable: On the Feasibility of Counterfeit Watermarks in Backdoor-Based Dataset Ownership Verification}

\author{Zhiying Li\inst{1} \and
Zhi Liu\inst{1} \and
Dongjie Liu\inst{1} \and
Shengda Zhuo\inst{1} \and
Guanggang Geng\inst{1} \and
Zhaoxin Fan\inst{3} \and
Shanxiang Lyu\inst{1}\thanks{Corresponding author.} \and
Xiaobo Jin\inst{2}\thanks{Corresponding author.} \and
Jian Weng\inst{1}
}
\authorrunning{Zhiying Li, et al.}
\institute{College of Cyber Security, Jinan University, Guangzhou 511436, China \and
School of Advanced Technology, Xi’an Jiaotong-Liverpool University, Suzhou 215000, China \and
School of Artificial Intelligence, Beihang University, Beijing 100083, China
}
\maketitle              % typeset the header of the contribution
\begin{abstract}
Backdoor watermarking has emerged as the predominant approach for protecting public datasets, enabling dataset ownership verification (DOV) through embedded triggers that induce predefined model behaviors. While existing works assume that DOV results can serve as reliable evidence for copyright infringement claims, we argue that this assumption is fundamentally flawed. In this paper, we expose critical vulnerabilities in current backdoor watermarking schemes by demonstrating that attackers can forge watermarks that are statistically indistinguishable from the original ones, thereby evading infringement allegations. Specifically, we propose a Forged Watermark Generator (FW-Gen), a lightweight variational autoencoder-based framework that generates forged watermarks preserving the statistical properties of original watermarks while exhibiting distinct visual patterns. Our attack operates under a realistic threat model where an accused attacker, upon receiving an infringement claim, extracts watermark information from the protected dataset and produces counterfeit evidence to refute the allegation. Extensive experiments across six backdoor watermarking methods, two benchmark datasets, and two model architectures demonstrate that forged watermarks achieve equivalent or superior statistical significance in hypothesis testing compared to original watermarks. These findings reveal that current DOV mechanisms are insufficient as standalone evidence for copyright disputes and call for more robust dataset protection schemes.

\end{abstract}

\section{Introduction}
\label{sec:introduction}

The proliferation of large-scale artificial intelligence models, such as GPT~\cite{mann2020language} and LLaMA~\cite{touvron2023llama}, underscores the critical role of high-quality datasets in modern machine learning. However, curating such datasets is a resource-intensive process involving meticulous data collection, annotation, and cleaning~\cite{krizhevsky2009learning,sakaridis2018semantic}. This substantial investment renders public datasets attractive targets for unauthorized use, where unethical actors may exploit these resources for commercial gain without compensating the original creators. Such practices not only violate intellectual property rights but also discourage data sharing within the open-source community, ultimately stifling innovation.

\begin{figure}[t]
    \centering
    \includegraphics[width=0.95\textwidth]{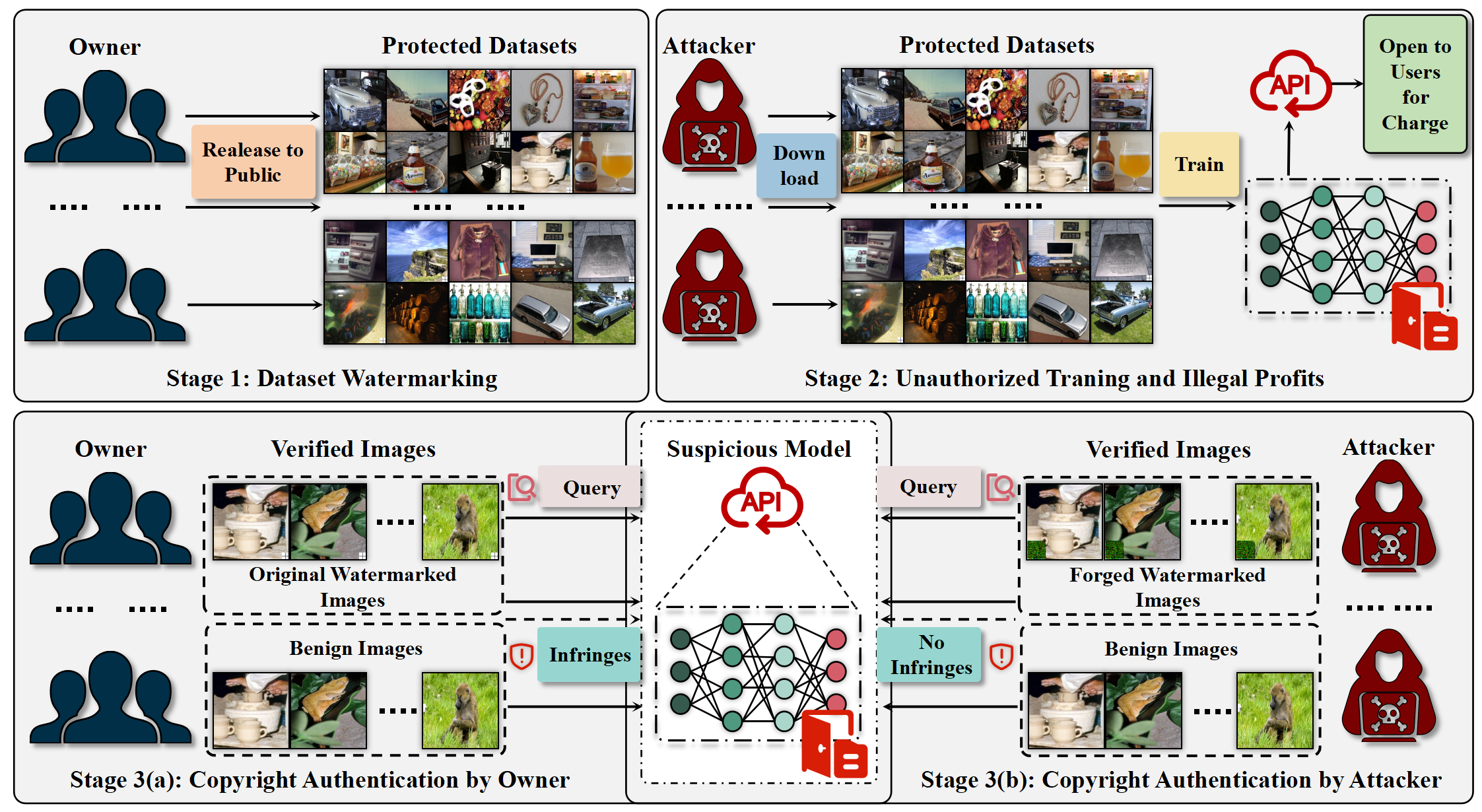}
%     \caption{Overview of backdoor watermarking. \textbf{Stage 1}: The dataset owner releases a watermarked dataset with target labels to the public (protected public dataset).
% \textbf{Stage 2}: The attacker uses the protected dataset to train a deep learning network without permission and provides a model API for profits.
% \textbf{Stage 3(a)}: The dataset owner queries the model API for responses to the original watermarked images and benign images, and determines copyright infringement.
% \textbf{Stage 3(b)}: The attacker queries the model API for responses to the forged watermarked images and benign images, and determines no copyright infringement.
% }
\caption{Overview of backdoor watermarking and our forgery attack. The dataset owner verifies ownership via original watermarks (Stage 3a), while the attacker evades detection using forged watermarks (Stage 3b).}
    \label{fig:backdoor}
\end{figure}

To mitigate these concerns, various dataset protection mechanisms have been proposed. Techniques designed for private datasets, including encryption~\cite{deng2020identity}, digital watermarking~\cite{guan2022deepmih}, and differential privacy~\cite{wang2021dpgen}, typically impose access restrictions or require control over the training process, rendering them unsuitable for public dissemination. In contrast, \emph{backdoor watermarking} has gained traction as a promising solution for public datasets, as it does not compromise data accessibility. This approach reframes dataset protection as a dataset ownership verification (DOV) problem~\cite{li2023black,guo2024domain}: before releasing a dataset, the owner embeds a trigger pattern into a subset of samples and later verifies ownership by checking whether a suspicious model exhibits the predefined backdoor behavior (Fig.~\ref{fig:backdoor}).

Despite its widespread adoption, a critical question remains unaddressed: \emph{Can DOV results alone reliably substantiate copyright infringement claims?} We contend that the answer is negative, based on two pivotal observations:

\begin{enumerate}
    \item \textbf{Lack of Temporal Binding.}  Existing schemes implicitly assume the dataset owner can prove temporal precedence of their watermark. In practice, however, timestamping mechanisms (e.g., blockchain registration~\cite{waheed2024grove}) are seldom employed due to prohibitive costs and operational complexity~\cite{zheng2018blockchain}. Most datasets are distributed via simple URL links, leaving no immutable record of watermark creation time.
    
    \item \textbf{Unrealistic Adversarial Assumptions.} Current methods presume that accused parties will passively accept infringement rulings. In reality, defendants in copyright disputes are strongly incentivized to gather counter-evidence to protect their commercial interests. If an adversary can produce a different watermark that elicits identical model behavior, the owner's claim becomes legally contestable.
\end{enumerate}

In this work, we systematically investigate the feasibility of \emph{forgery attacks} against backdoor watermarking schemes. We consider a realistic scenario wherein an accused attacker, upon receiving an infringement allegation, extracts the original watermark from the protected dataset and forges a visually distinct yet statistically equivalent counterpart. By presenting this counterfeit as evidence, the attacker can cast reasonable doubt on the uniqueness of the owner's watermark, thereby undermining the legal standing of the DOV outcome.

Our contributions are fourfold:
\begin{itemize}
    \item We identify and formalize two fundamental limitations of current backdoor watermarking schemes that enable forgery attacks: the absence of temporal binding and the behavioral equivalence of distinct triggers.
    
    \item We propose \emph{FW-Gen}, a lightweight variational autoencoder-based framework that generates forged watermarks preserving the statistical properties of originals while ensuring visual dissimilarity.
    
    \item We formally prove that behavioral-verification-only schemes are inherently vulnerable (Theorem 1), establishing necessary conditions for forgery-resistant DOV.
    
    \item Through extensive experiments across six watermarking methods, two datasets, and two model architectures, we demonstrate that forged watermarks consistently achieve equivalent or superior statistical significance in DOV, highlighting the inadequacy of current schemes as standalone evidence.
\end{itemize}

\textbf{Ethical Considerations.} We recognize the dual-use nature of this research. While our attack could be misused, we believe responsible disclosure is essential for advancing the security of dataset protection mechanisms. We adhere to established disclosure practices and will release our code to facilitate the development of robust defenses.

\section{Related Work}
\label{sec:related}
Below we review related works on backdoor attacks and dataset protection.

\subsection{Backdoor Attacks}

Backdoor attacks embed hidden malicious behaviors into machine learning models during training. Poisoned models perform normally on clean inputs but produce attacker-specified outputs when presented with trigger patterns. BadNets~\cite{gu2017badnets} pioneered this area by demonstrating that simple pixel-patch triggers can effectively compromise deep neural networks. Subsequent work has explored diverse trigger designs: Blended attacks~\cite{chen2017targeted} use transparency-based trigger embedding for improved stealth; Trojan attacks~\cite{liu2018trojaning} optimize triggers to maximize activation in specific neurons; and steganographic approaches~\cite{li2020invisible} employ regularization techniques to create imperceptible triggers.

Recent research has focused on enhancing attack stealthiness to evade detection. Latent backdoors~\cite{doan2021backdoor} craft triggers based on anomalies in the model's latent representations. Adaptive perturbation techniques~\cite{zhao2022defeat} dynamically adjust trigger patterns to circumvent defense mechanisms. Clean-label attacks~\cite{tang2023did} maintain correct labels for poisoned samples, making detection more challenging.

A common paradigm across existing backdoor attacks is that the triggers used during training are identical to those used during inference. Our work challenges this assumption by demonstrating that different triggers can induce equivalent backdoor behaviors.

\subsection{Dataset Protection}

Dataset protection mechanisms can be categorized based on the target data accessibility.

\noindent \textbf{Private Dataset Protection.} Encryption~\cite{deng2020identity,wang2016efficient} protects sensitive data by restricting access to authorized users possessing decryption keys. Digital watermarking~\cite{guan2022deepmih,bansal2022certified} embeds ownership information into data without revealing details to unauthorized parties. Differential privacy~\cite{wang2021dpgen} protects individual sample membership by introducing controlled noise during model training. These methods require either access control or training process supervision, making them unsuitable for public datasets.

\noindent \textbf{Public Dataset Protection.} Backdoor watermarking reframes public dataset protection as a DOV problem~\cite{li2023black}. Dataset owners embed trigger patterns into a subset of samples before publication and later verify ownership by testing suspicious models for predefined backdoor behaviors. Poison-label approaches~\cite{li2023black} modify sample labels to target classes, while clean-label methods~\cite{tang2023did} preserve original labels through more sophisticated trigger designs. Domain generalization techniques~\cite{guo2024domain} enhance watermark robustness by enabling correct classification of challenging samples.

Existing backdoor watermarking relies on the consistency of backdoor attacks for copyright authentication, and implicitly make a strong assumption that the attacker will not refute the accusation. Inspired by these findings, we rethink the copyright authentication process from the perspective of forgery attacks, revealing significant security vulnerabilities.

\subsection{Attacks on Watermarking Methods}

The security of watermarking schemes has been challenged from multiple angles. In model watermarking, ambiguity attacks~\cite{fan2019rethinking} demonstrate that adversaries can embed competing watermarks, creating ownership disputes. Removal attacks attempt to eliminate watermarks through fine-tuning, pruning, or knowledge distillation while preserving model utility. For image watermarking, copy attacks and protocol attacks have exposed similar vulnerabilities.

In the context of dataset watermarking, the threat landscape differs fundamentally. Unlike model watermarks embedded in parameters, backdoor watermarks manifest as learned behavioral patterns. This distinction opens a new attack surface: rather than removing or overwriting watermarks, an attacker can forge alternative watermarks inducing equivalent behaviors. To our knowledge, this forgery attack vector has not been explored in prior work on DOV, representing a significant gap in security analysis.

%==============================================================================
% PRELIMINARIES
%==============================================================================
\section{Preliminaries and Theoretical Foundation}
\label{sec:preliminaries}

% In this section, establishes the formal framework for backdoor watermarking and dataset ownership verification, followed by our theoretical analysis of forgery vulnerabilities.

In this section, we will provide the processes including backdoor watermarking, dataset ownership verification with hypothesis testing, treat model and theoretical analysis of forgery vulnerabilities.

\subsection{Backdoor Watermarking for Dataset Ownership Verification}

We introduce backdoor watermarking in the field of image classification. Let $D = \{(x_i, y_i)\}_{i=1}^{N}$ denote a clean dataset, where $x_i \in [0, 255]^{C \times W \times H}$ represents the $i$-th image and $y_i \in \{0, 1, \ldots, K-1\}$ is its corresponding ground-truth label.

\begin{definition}[Backdoor Watermarking]
\label{def:watermarking}
A backdoor watermarking scheme $\mathcal{W} = (\mathsf{Embed}, \mathsf{Verify})$ consists of two algorithms:
\begin{itemize}
    \item $\mathsf{Embed}(D, t, \yhat, \lambda) \rightarrow \Dp$: Given a clean dataset $D$, trigger pattern $t$, target label $\yhat$, and watermark rate $\lambda$, outputs a protected dataset $\Dp$ where a fraction $\lambda$ of samples are watermarked.
    \item $\mathsf{Verify}(\ftilde, t, \yhat) \rightarrow \{0, 1\}$: Given a suspicious model $\ftilde$, trigger pattern $t$, and target label $\yhat$, outputs 1 if the model exhibits the expected backdoor behavior, and 0 otherwise.
\end{itemize}
\end{definition}

The embedding process selects $\lambda \cdot N$ samples uniformly at random, applies the trigger function $\mathcal{G}(x_i, t)$ to embed the watermark pattern, and modifies their labels to $\yhat$. The resulting protected dataset is:
\begin{equation}
\Dp = \{(x_i, y_i) : i \notin S\} \cup \{(\mathcal{G}(x_i, t), \yhat) : i \in S\}
\end{equation}
where $S$ denotes the set of selected sample indices with $|S| = \lambda \cdot N$.

\subsection{Dataset Ownership Verification via Hypothesis Testing} \label{sec:dov}

DOV determines whether a suspicious model $\ftilde$ was trained on the protected dataset $\Dp$ by testing for backdoor behavior. Two verification scenarios are considered:
\begin{itemize}
    \item \textbf{Stealing Model Scenario (S).} The suspicious model $\ftilde$ is trained on $\Dp$ and should exhibit backdoor behavior---classifying watermarked inputs to the target label $\yhat$.
    \item \textbf{Independent Model Scenario (I).} A benign model $f$ is trained only on clean data and should not exhibit backdoor behavior---classifying watermarked inputs according to their original semantic content.
\end{itemize}

Depending on the API type, different statistical tests are employed:
\begin{itemize}
    \item \textbf{Probability-Available API.} When the model outputs class probability distributions $p \in [0,1]^K$, a paired t-test compares the probabilities assigned to the target class for watermarked versus benign images. The null hypothesis $H_0$ states that watermarked and benign images yield identical probability distributions.
    \item \textbf{Label-Only API.} When only predicted labels $c \in \{0, 1, \ldots, K-1\}$ are available, the Wilcoxon signed-rank test compares label predictions. The null hypothesis $H_0$ states that watermarked images are not preferentially classified as the target label.
\end{itemize}

Rejecting $H_0$ at significance level $\alpha$ (typically 0.05) indicates that the suspicious model was likely trained on the protected dataset.

\subsection{Threat Model}
\label{sec:threat}

We formalize the threat model by specifying the goals, capabilities, and knowledge of both parties in a DOV dispute.

\noindent \textbf{Parties.} We consider two parties: (1) the \emph{defender} (dataset owner) who releases a watermarked public dataset and later claims copyright infringement, and (2) the \emph{attacker} (model trainer) who is accused of unauthorized dataset usage and seeks to refute the allegation.

\noindent \textbf{Defender's Goal and Capabilities.} The defender aims to prove that the attacker's model was trained on the protected dataset $\Dp$. The defender has complete control over the dataset before release, including the ability to embed arbitrary watermarks. After the attacker deploys a model, the defender can only interact with it through black-box API queries, without access to model parameters or training data.

\noindent \textbf{Attacker's Goal and Capabilities.} The attacker's goal is to evade the infringement allegation by producing counter-evidence that casts reasonable doubt on the defender's claim. The attacker has full control over their deployed model $\ftilde$ and unrestricted access to the published dataset $\Dp$. The attacker can train additional models on data extracted from $\Dp$.

\noindent \textbf{Attacker's Knowledge.} We assume the attacker becomes aware of the watermark's existence only after receiving the infringement claim. Subsequently, the attacker can:
\begin{itemize}
    \item[\textbf{(K1)}] Detect and extract watermarked samples from $\Dp$ using existing backdoor detection methods~\cite{peri2020deep,zeng2021rethinking};
    \item[\textbf{(K2)}] Infer the target label $\yhat$ by querying their own model with detected watermarked samples;
    \item[\textbf{(K3)}] Train a benign model $f$ on clean data (\ie, $\Dp$ with watermarked samples removed).
\end{itemize}

\noindent \textbf{Justification.} Assumption (K1) is empirically supported by our experiments in Section~\ref{sec:detection}, where frequency-domain detection achieves near-perfect accuracy (>99\%) for most watermarking methods. Assumption (K2) follows trivially from (K1) and model access. Assumption (K3) is realistic since the attacker possesses computational resources sufficient to train the suspicious model.

\subsection{Theoretical Analysis of Forgery Vulnerability}
\label{sec:theory}

We establish the theoretical foundation for forgery attacks by formalizing the conditions under which alternative watermarks can evade DOV.

\begin{definition}[Behavioral Equivalence]
\label{def:equivalence}
Two watermarks $t_1$ and $t_2$ are behaviorally equivalent with respect to model $\ftilde$ and target label $\yhat$, denoted $t_1 \sim_{\ftilde, \yhat} t_2$, if for any input $x$:
\begin{equation}
\Pr[\ftilde(\mathcal{G}(x, t_1)) = \yhat] = \Pr[\ftilde(\mathcal{G}(x, t_2)) = \yhat]
\end{equation}
\end{definition}

\begin{definition}[Successful Forgery]
\label{def:forgery}
A forgery attack is successful if it produces a watermark $\tfw$ such that:
\begin{enumerate}
    \item \textbf{Behavioral Equivalence:} $\tfw \sim_{\ftilde, \yhat} \tow$ where $\tow$ is the original watermark;
    \item \textbf{Visual Distinctiveness:} $d(\tfw, \tow) > \tau$ for some distance metric $d$ and threshold $\tau$;
    \item \textbf{Clean Model Consistency:} For any benign model $f$, $\Pr[f(\mathcal{G}(x, \tfw)) = y_x] \approx \Pr[f(\mathcal{G}(x, \tow)) = y_x]$ where $y_x$ is the true label of $x$.
\end{enumerate}
\end{definition}

\begin{theorem}[Forgery Vulnerability]
\label{thm:vulnerability}
Any backdoor watermarking scheme $\mathcal{W}$ that relies solely on behavioral verification (\ie, $\mathsf{Verify}$ only tests model responses to watermarked inputs) is vulnerable to forgery attacks. Specifically, if an attacker can find a watermark $\tfw$ satisfying Definition~\ref{def:forgery}, then the DOV result for $\tfw$ is statistically indistinguishable from that for $\tow$.
\end{theorem}

\begin{proof}
The verification algorithm $\mathsf{Verify}$ computes a test statistic $T$ based on model responses:
\begin{equation}
T = g(\{\ftilde(\mathcal{G}(x_i, t))\}_{i=1}^{n})
\end{equation}
for some function $g$ (\eg, Paired T-test statistic or Wilcoxon Signed-Rank Test).

By behavioral equivalence (Definition~\ref{def:forgery}, condition 1), the random variables $\ftilde(\mathcal{G}(X, \tfw))$ and $\ftilde(\mathcal{G}(X, \tow))$ have identical distributions for $X$ drawn from the data distribution. Since $T$ is a function of i.i.d. samples from these distributions, by the continuous mapping theorem: $T_{\text{fw}} \stackrel{d}{=} T_{\text{ow}}$.

Consequently, the p-values satisfy $p_{\text{fw}} = p_{\text{ow}}$, and $\mathsf{Verify}$ produces identical decisions for both watermarks. By visual distinctiveness, $\tfw \neq \tow$, allowing the attacker to present $\tfw$ as independent evidence. Without temporal binding, the defender cannot prove that $\tow$ preceded $\tfw$, rendering the DOV result ambiguous.
\end{proof}

\begin{corollary}
To achieve forgery-resistant DOV, watermarking schemes must incorporate mechanisms beyond behavioral verification, such as cryptographic timestamping or watermark binding to verifiable external records.
\end{corollary}

%==============================================================================
% METHODOLOGY
%==============================================================================
\begin{figure*}[t ]
    \centering
    \includegraphics[width=0.95\textwidth]{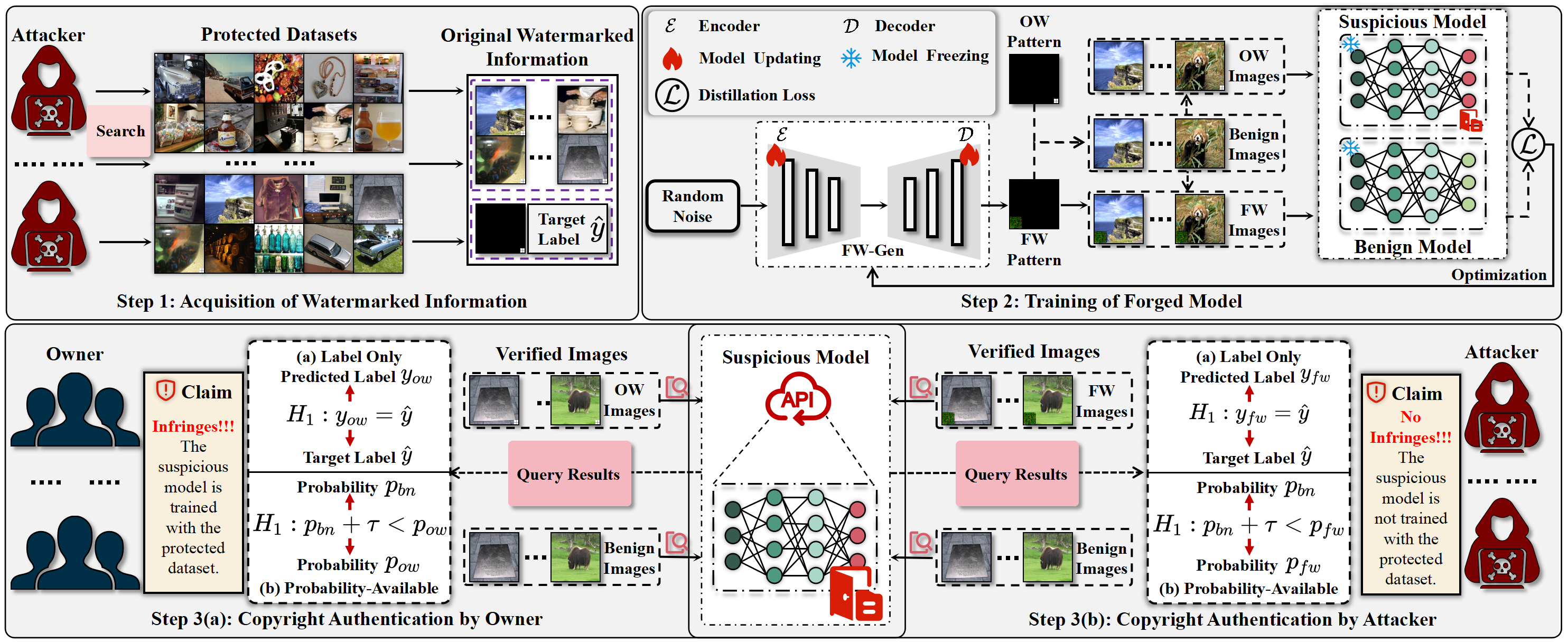}
    % \caption{The proposed forgery attack scheme is summarized, including several key steps: acquisition of watermark information, training of forged model, and copyright authentication by the owner and the attacker, where OW represents the original watermark and FW represents the forged watermark. $p_{ow}$ and $y_{ow}$ denote the predicted probability and predicted label of original watermarked images when querying the suspicious model. $p_{fw}$ and $y_{fw}$ denote the predicted probability and predicted label of forged watermarked images when querying the suspicious model. $\hat{y}$ represents the target label. $\tau$ is a predefined tolerance threshold.}
    \caption{FW-Gen pipeline: \textbf{Step 1} watermark extraction from $D_p$; \textbf{Step 2} forged watermark generation via VAE; and \textbf{Step 3} ownership dispute using forged evidence. OW: original watermark; FW: forged watermark.}
    \label{fig:forgery}
\end{figure*}

\section{Methodology}
  \label{sec:methodology}

This section details the forgery attack pipeline illustrated in Fig.~\ref{fig:forgery}. Given a protected dataset $D_p$ with embedded watermarks, the attacker aims to generate forged watermarks that achieve statistical equivalence with original watermarks in DOV while maintaining visual distinctiveness.

\subsection{Watermark Information Extraction}
\label{subsec:extraction}

Following the threat model in Section~\ref{sec:threat}, the attacker extracts watermarked samples from $D_p$ using frequency-domain analysis~\cite{zeng2021rethinking}, which detects anomalous high-frequency signals characteristic of backdoor-embedded images. The target label $\hat{y}$ is subsequently inferred by querying the suspicious model $\tilde{f}$ with detected watermarked samples. As validated in Section~\ref{sec:detection}, this detection achieves $>$99\% accuracy for most watermarking methods, confirming the practical feasibility of our threat model assumptions.

\subsection{Forged Watermark Generation}
\label{subsec:generation}

\textbf{Network Architecture.}
We employ a lightweight Variational Autoencoder (VAE) to generate forged watermarks that preserve statistical properties while ensuring visual distinctiveness. Unlike standard VAEs that reconstruct inputs, FW-Gen takes random noise as input to guarantee the generated watermark differs visually from the original:
\begin{equation}
      t_{fw} = \mathcal{D}_\varrho(\mu + \sigma \odot \epsilon), \quad \text{where} \ (\mu, \sigma) = \mathcal{E}_\phi(\epsilon'), \ \epsilon, \epsilon' \sim \mathcal{N}(0, I)
      \label{eq:vae}
\end{equation}
where $\mathcal{E}_\phi$ and $\mathcal{D}_\varrho$ denote the encoder and decoder parameterized by $\phi$ and $\varrho$, respectively. Both components consist of three convolutional blocks, minimizing computational overhead while maintaining generation quality.

\noindent \textbf{Training Objective.}
FW-Gen is trained using both the suspicious model $\tilde{f}$ and a benign model $f$ (trained on clean data extracted from $D_p$). Inspired by knowledge distillation~\cite{hinton2015distilling}, we design dual losses to transfer behavioral characteristics from original to forged watermarks:
\begin{equation}
      \mathcal{L} = \mathcal{L}_B + \mathcal{L}_W
      \label{eq:total_loss}
\end{equation}

The benign model loss $\mathcal{L}_B$ ensures forged watermarks do not introduce detectable artifacts on models that have not been exposed to the watermark during training:
% \begin{equation}
%       \mathcal{L}_B = \sum_{i=1}^{N} \alpha T^2 \cdot \text{CE}(p(\mathcal{G}(x_i, t_{ow})), p(\mathcal{G}(x_i, t_{fw}))) \\
%       + (1-\alpha) \cdot \text{CE}(\text{one-hot}(y_i), p(\mathcal{G}(x_i, t_{fw})))
%       \label{eq:loss_benign}
% \end{equation}
\begin{equation}
\begin{aligned}
\mathcal{L}_B
&= \sum_{i=1}^{N} \alpha T^2 \cdot
\text{CE}\!\left(
p(\mathcal{G}(x_i, t_{ow})),
p(\mathcal{G}(x_i, t_{fw}))
\right) \\
&\quad + (1-\alpha) \cdot
\text{CE}\!\left(
\text{one-hot}(y_i),
p(\mathcal{G}(x_i, t_{fw}))
\right)
\end{aligned}
\label{eq:loss_benign}
\end{equation}

The suspicious model loss $\mathcal{L}_W$ aligns the backdoor behavior triggered by forged watermarks with that of original watermarks:
% \begin{equation}
%       \mathcal{L}_W = \sum_{i=1}^{N} \alpha T^2 \cdot \text{CE}(\tilde{p}(\mathcal{G}(x_i, t_{ow})), \tilde{p}(\mathcal{G}(x_i, t_{fw}))) + (1-\alpha) \cdot \text{CE}(\text{one-hot}(\hat{y}), \tilde{p}(\mathcal{G}(x_i, t_{fw})))
%       \label{eq:loss_watermark}
% \end{equation}
\begin{equation}
\begin{aligned}
\mathcal{L}_W
&= \sum_{i=1}^{N} \alpha T^2 \cdot
\text{CE}\!\left(
\tilde{p}(\mathcal{G}(x_i, t_{ow})),
\tilde{p}(\mathcal{G}(x_i, t_{fw}))
\right) \\
&\quad + (1-\alpha) \cdot
\text{CE}\!\left(
\text{one-hot}(\hat{y}),
\tilde{p}(\mathcal{G}(x_i, t_{fw}))
\right)
\end{aligned}
\label{eq:loss_watermark}
\end{equation}
where $T$ is the temperature parameter, $\alpha$ is a weighting coefficient, $\text{CE}(\cdot, \cdot)$ denotes cross-entropy, and $p(\cdot)$, $\tilde{p}(\cdot)$ represent the probability outputs of the benign and suspicious models, respectively. The function $\mathcal{G}(x, t)$ applies trigger pattern $t$ to image $x$, and $\text{one-hot}(\cdot)$ converts a label to its one-hot encoding.

 The dual losses serve complementary purposes: $\mathcal{L}_B$ prevents the forged watermark from being distinguishable on clean models, while $\mathcal{L}_W$ ensures equivalent backdoor activation on the suspicious model. Both components are essential for successful forgery.

\subsection{Ownership Dispute via Forged Evidence}
\label{subsec:dispute}

After training, the attacker generates forged watermarks and performs DOV following the hypothesis testing procedure in Section~\ref{sec:dov}. The attacker queries the suspicious model $\tilde{f}$ with both forged watermarked images $x_{fw} = \mathcal{G}(x, t_{fw})$ and benign images $x$, then conducts the same statistical tests used by the dataset owner.

If the forged watermark achieves equivalent statistical significance---rejecting $H_0$ in the stealing model scenario and accepting $H_0$ in the independent model scenario---the attacker can present it as counter-evidence to dispute the owner's infringement claim. Since the defender cannot prove temporal precedence of their watermark without cryptographic timestamping mechanisms, the DOV result becomes legally ambiguous, effectively undermining the evidentiary value of backdoor watermarking for copyright disputes.

%==============================================================================
% EXPERIMENTS
%==============================================================================
\section{Experiments}
\label{sec:experiments}

% We conduct comprehensive experiments to evaluate FW-Gen across diverse settings. Our evaluation addresses three research questions:
% \begin{itemize}
%     \item \textbf{RQ1:} Can attackers reliably extract watermark information from protected datasets?
%     \item \textbf{RQ2:} Do forged watermarks achieve statistical equivalence with original watermarks in DOV?
%     \item \textbf{RQ3:} How do individual loss components contribute to forgery effectiveness?
% \end{itemize}
We conduct comprehensive experiments to answer three research questions:

\noindent \textbf{RQ1:} Can attackers reliably extract watermark information from protected datasets?

\noindent \textbf{RQ2:} Do forged watermarks achieve statistical equivalence with original watermarks in DOV?

\noindent \textbf{RQ3:} How do individual loss components contribute to forgery effectiveness?

\subsection{Experimental Setup}

\noindent \textbf{Datasets and Models.}
We evaluate on CIFAR-10(C10)~\cite{krizhevsky2009learning} (60K images, $32 \times 32$) and a 100K subset of ImageNet(IM)~\cite{deng2009imagenet} (resized to $224 \times 224$). Classification models include ResNet-18 (Res)~\cite{he2016deep} and VGG-19 (VGG)~\cite{simonyan2014very}.

\noindent \textbf{Watermarking Methods.}
We consider six representative backdoor watermarking approaches: BadNets~\cite{gu2017badnets} (pixel-patch triggers with Cross and Line patterns), Blended~\cite{chen2017targeted} (transparency-based embedding), $\ell_0$-invisible~\cite{li2020invisible} (steganographic triggers), Nature~\cite{chen2017targeted} (natural image blending), Trojan-sq~\cite{liu2018trojaning} (optimized square triggers), and Trojan-wm~\cite{liu2018trojaning} (optimized watermark-style triggers).

\noindent \textbf{Implementation.}
Watermark rate $\lambda = 0.1$. FW-Gen uses temperature $T \in \{500, 800\}$, learning rate 0.008, 500 iterations, batch size 32, and AdamW optimizer. All experiments run on a single NVIDIA L20 GPU.

\noindent \textbf{Metrics.}
We measure: (1) \textbf{Statistical equivalence} via $p$-value and $\Delta P$ (probability difference); (2) \textbf{Classification performance} via benign accuracy (BA), original/forged watermark success rates (OWSR/FWSR); (3) \textbf{Visual distinctiveness} via peak signal-to-noise ratio (PSNR)\cite{wang2019spatial}, mean square error (MSE) and structural similarity index measure (SSIM)\cite{wang2004image}.

\subsection{Watermark Detection Effectiveness (RQ1)}
\label{sec:detection}

We validate that attackers can reliably detect watermarked samples using frequency-domain analysis~\cite{zeng2021rethinking}. Table~\ref{tab:detection} presents detection accuracy across watermarking methods on CIFAR-10.

\begin{table}[t]
\centering
\caption{Watermark detection accuracy (\%) on CIFAR-10.}
\label{tab:detection}
\begin{tabular}{lcccccc}
\toprule
Method & BadNets & Blended & $\ell_0$-inv & Nature & Trojan-sq & Trojan-wm \\
\midrule
Accuracy & 90.2 & 99.1 & 100 & 99.4 & 99.8 & 99.9 \\
\bottomrule
\end{tabular}
\end{table}

All methods except BadNets achieve near-perfect detection accuracy (>99\%), confirming that watermarked samples can be reliably extracted in black-box settings. BadNets' lower accuracy (90.2\%) stems from its simpler trigger patterns, but 90\% accuracy is sufficient for practical forgery attacks.

%========================================================
\begin{table}[t]

% \caption{Comparison of the statistical significance of forged watermarks and original watermarks for various backdoor watermarking methods in a probabilistic scenario on the CIFAR-10 (C10) and ImageNet (IM) datasets, where M and BW represent the classification model and the backdoor watermarking method, respectively. Badnets and Blended methods consider two different shapes: cross and line, respectively. Red marks indicate that the forged watermarks show stronger statistical significance than the original watermarks.}
\caption{Statistical significance comparison (probability-available API) on CIFAR-10 and ImageNet. Format: original $|$ forged. \textcolor{red}{Red}: forged outperforms original.}
\centering
% \begin{subtable}
\resizebox{0.95\textwidth}{!}{%

\begin{tabular}{c|c|c|c|c|c|c|c|c|c|c}

\hline
     &  & BW & \multicolumn{4}{c|}{Badnets} & \multicolumn{4}{c}{Blended} \\
\cline{3-11}
     &  & Shape & \multicolumn{2}{c|}{Cross}  & \multicolumn{2}{c|}{Line} & \multicolumn{2}{c|}{Cross} & \multicolumn{2}{c}{Line}  \\
\cline{3-11}
    \multirow{-3}{*}{Data}  & \multirow{-3}{*}{M}  & Metric & $p$-value & $\Delta P$ & $p$-value  & $\Delta P$ & $p$-value & $\Delta P$  & $p$-value & $\Delta P$  \\
\hline
    &   & S & 1.2e${-173}$ $|$ \color{red}{1.1e${-173}$} & 9.9e${-1}$ $|$ 9.9e${-1}$                       & 2.1e${-107}$ $|$ 1.8e${-101}$ & 9.9e${-1}$ $|$ 9.8e${-1}$   & 
    
    4.8e${-164}$ $|$ \color{red}{9.5e${-165}$} & 9.9e${-1}$ $|$ 9.9e${-1}$ & 5.0e${-54}$ $|$ \color{red}{1.4e${-121}$} & 9.3e${-1}$ $|$ \color{red}{9.8e${-1}$}   \\
\cline{3-11}
    & \multirow{-2}{*}{Res} & I  & 1 $\quad |\quad$ 1   & 7.9e${-4}$ $|$ 7.9e${-4}$ & 1 $\quad |\quad$ 1 & -3.6e${-4}$ $|$ -3.6e${-4}$ & 
    1 $\quad |\quad$ 1 & 6.2e${-1}$ $|$ 6.2e${-1}$ & 1 $\quad |\quad$ 1  & -2.3e${-4}$ $|$ -2.3e${-4}$ \\
\cline{2-11}
    &  & S & 7.3e${-144}$ $|$ 1.1e${-143}$ & 9.9e${-1}$ $|$ 9.9e${-1}$ & 4.6e${-92}$ $|$ 3.2e${-75}$   & 9.8e${-1}$ $|$ 9.7e${-1}$  
    
    & 6.8e${-241}$ $|$ \color{red}{3.7e${-241}$ }                     & 9.9e${-1}$ $|$ 9.9e${-1}$ & 7.7e${-59}$ $|$ \color{red}{1.5e${-96}$} & 9.4e${-1}$ $|$ \color{red}{9.7e${-1}$}   \\
\cline{3-11}
    \multirow{-4}{*}{C10} & \multirow{-2}{*}{VGG}    & I  & 1 $\quad |\quad$ 1  & 5.4e${-5}$ $|$ 5.4e${-5}$ & 1 $\quad |\quad$ 1  & -5.8e${-5}$ $|$ -5.8e${-5}$  
    
    & 1 $\quad |\quad$ 1 & 1.8e${-4}$ $|$1.8e${-4}$ & 1 $\quad |\quad$ 1      & -1.9e${-4}$ $|$ -1.9e${-4}$ \\
\hline

    &   & S & 6.2e${-180}$ $|$ 1.4e${-177}$ & 9.9e${-1}$ $|$ 9.9e${-1}$                       & 8.6e${-59}$ $|$ \color{red}{4.3e${-63}$} & 9.4e${-1}$ $|$ 9.3e${-1}$   & 
    
    1.6e${-30}$ $|$ \color{red}{1.3e${-65}$} & 7.6e${-1}$ $|$ \color{red}{8.8e${-1}$} & 2.2e${-18}$ $|$ \color{red}{2.4e${-25}$} & 6.5e${-1}$ $|$ 6.5e${-1}$   \\
\cline{3-11}
    & \multirow{-2}{*}{Res} & I  & 1 $\quad |\quad$ 1   & 8.2e${-6}$ $|$ 8.2e${-6}$ & 1 $\quad |\quad$ 1 & 5.2e${-6}$ $|$ 5.2e${-6}$ & 
    
    1 $\quad |\quad$ 1 & 5.0e${-6}$ $|$ 5.0e${-6}$ & 1 $\quad |\quad$ 1  & -2.2e${-7}$ $|$ -2.2e${-7}$ \\
\cline{2-11}
    &  & S & 2.2e${-181}$ $|$ 2.2e${-181}$ & 9.9e${-1}$ $|$ 9.9e${-1}$ & 9.3e${-62}$ $|$ \color{red}{3.8e${-112}$}   & 9.5e${-1}$ $|$ \color{red}{9.7e${-1}$}  & 
    
    8.9e${-49}$ $|$ \color{red}{2.1e${-77}$}  & 8.8e${-1}$ $|$ \color{red}{9.5e${-1}$} & 1.2e${-31}$ $|$ \color{red}{7.6e${-72}$} & 8.1e${-1}$ $|$ \color{red}{9.5e${-1}$}   \\
\cline{3-11}
    \multirow{-4}{*}{IM} & \multirow{-2}{*}{VGG}    & I  & 1 $\quad |\quad$ 1  & -1.9e${-7}$ $|$ -1.9e${-7}$ & 1 $\quad |\quad$ 1  & 2.7e${-6}$ $|$ 2.7e${-6}$  
    & 
    1 $\quad |\quad$ 1 & 6.7e${-8}$ $|$ 6.7e${-8}$ & 1 $\quad |\quad$ 1      & 2.7e${-6}$ $|$ 2.7e${-6}$ \\
% \bottomrule[1pt]
\hline
\end{tabular}}
%\caption{The comparison results (Original Watermark $|$ Forged Watermark) of Badnets and Blended, with Cross and Line denoting different style patterns of original watermarks.}
\label{ttest1}
% \end{subtable}

\vspace{1pt}

% \begin{subtable}
\centering
\resizebox{0.95\textwidth}{!}{

\begin{tabular}{c|c|c|cc|cc|cc|cc}

\hline
      &   & BW &\multicolumn{2}{c|}{$l_0$ inv}    & \multicolumn{2}{c|}{Nature}     & \multicolumn{2}{c|}{Trojan\_sq}     & \multicolumn{2}{c}{Trojan\_wm}                                                                           \\ \cline{3-11} 
\multirow{-2}{*}{Data}  & \multirow{-2}{*}{M}   & Metric & \multicolumn{1}{c|}{$p$-value}      & $\Delta P$    & \multicolumn{1}{c|}{$p$-value}     & $\Delta P$    & \multicolumn{1}{c|}{$p$-value}      & $\Delta P$  & \multicolumn{1}{c|}{$p$-value}   & $\Delta P$   \\ \hline
    &  & S   & \multicolumn{1}{c|}{7.7e${-215}$ $|$ 4.1e${-214}$} & 9.9e${-1}$ $|$ 9.9e${-1}$   & \multicolumn{1}{c|}{2.7e${-146}$ $|$ \color{red}{2.3e${-146}$}} & 9.9e${-1}$ $|$ 9.9e${-1}$ & \multicolumn{1}{c|}{2.0e${-247}$ $|$ 2.2e${-247}$} & 9.9e${-1}$ $|$ 9.9e${-1}$ & \multicolumn{1}{c|}{4.6e${-192}$ $|$ \color{red}{3.2e${-192}$}} & 9.9e${-1}$ $|$ 9.9e${-1}$ \\ 
\cline{3-11} 
    & \multirow{-2}{*}{Res} & I   & \multicolumn{1}{c|}{1 $\quad |\quad$ 1}              & 2.5e${-4}$ $|$ 2.5e${-4}$   & \multicolumn{1}{c|}{1 $\quad |\quad$ 1} & 2.6e${-2}$ $|$ 2.6e${-2}$ & \multicolumn{1}{c|}{1 $\quad |\quad$ 1}   & 4.1e${-3}$ $|$ 4.1e${-3}$ & \multicolumn{1}{c|}{1 $\quad |\quad$ 1}                     & 1.7e${-2}$ $|$ 1.7e${-2}$ \\ 
\cline{2-11} 
    &  & S   & \multicolumn{1}{c|}{3.7e${-184}$ $|$ 8.8e${-115}$} & 2.5e${-4}$ $|$ 2.5e${-4}$   & \multicolumn{1}{c|}{3.5e${-256}$ $|$ 3.8e${-111}$} & 9.9e${-1}$ $|$ 9.8e${-1}$ & \multicolumn{1}{c|}{1.6e${-227}$ $|$ \color{red}{1.5e${-227}$}} & 9.9e${-1}$ $|$ 9.9e${-1}$ & \multicolumn{1}{c|}{ 1.5e${-124}$ $|$ 1.5e${-124}$} & 9.9e${-1}$ $|$ 9.9e${-1}$ \\ 
\cline{3-11} 
    \multirow{-4}{*}{C10} & \multirow{-2}{*}{VGG} & I   & \multicolumn{1}{c|}{1 $\quad |\quad$ 1}              & -5.3e${-3}$ $|$ -5.3e${-3}$ & \multicolumn{1}{c|}{1 $\quad |\quad$ 1}  & 5.2e${-2}$ $|$ 5.2e${-2}$ & \multicolumn{1}{c|}{1 $\quad |\quad$ 1} & 6.4e${-2}$ $|$ 6.4e${-2}$ & \multicolumn{1}{c|}{1 $\quad |\quad$ 1}     & 8.6e${-2}$ $|$ 8.6e${-2}$ \\ 
\hline
    \multicolumn{1}{l|}{}  &   & S   & \multicolumn{1}{c|}{1.0e${-240}$ $|$ \color{red}{1.3e${-241}$}} & 9.9e${-1}$ $|$ 9.9e${-1}$   & \multicolumn{1}{c|}{1.0e${-224}$ $|$ 4.5e${-181}$} & 9.9e${-1}$ $|$ 9.9e${-1}$ & \multicolumn{1}{c|}{6.1e${-199}$ $|$ 9.9e${-131}$} & 9.9e${-1}$ $|$ 9.8e${-1}$ & \multicolumn{1}{c|}{2.9e${-236}$ $|$ \color{red}{2.3e${-236}$}} & 9.9e${-1}$ $|$ 9.9e${-1}$ \\ 
\cline{3-11} 
    \multicolumn{1}{l|}{} & \multirow{-2}{*}{Res} & I   & \multicolumn{1}{c|}{1 $\quad |\quad$ 1} & 9.9e${-5}$ $|$ 9.9e${-5}$   & \multicolumn{1}{c|}{1 $\quad |\quad$ 1}   & 1.6e${-3}$ $|$ 1.6e${-3}$ & \multicolumn{1}{c|}{1 $\quad |\quad$ 1} & 1.0e${-3}$ $|$ 1.0e${-3}$ & \multicolumn{1}{c|}{1 $\quad |\quad$ 1}                & 2.1e${-2}$ $|$ 2.1e${-2}$ \\ 
\cline{2-11} 
    \multicolumn{1}{l|}{} &  & S   & \multicolumn{1}{c|}{0 $\quad |\quad$ 0} & 1 $\quad |\quad$ 1 & \multicolumn{1}{c|}{0 $\quad |\quad$ 0} & 1 $\quad |\quad$ 1  & \multicolumn{1}{c|}{0 $\quad |\quad$ 0} & 1 $\quad |\quad$ 1          & \multicolumn{1}{c|}{0 $\quad |\quad$ 0} & 1 $\quad |\quad$ 1  \\ 
\cline{3-11} 
    \multicolumn{1}{l|}{\multirow{-4}{*}{IM}} & \multirow{-2}{*}{VGG} & I   & \multicolumn{1}{c|}{1 $\quad |\quad$ 1} & 3.6e${-6}$ $|$ 3.6e${-6}$   & \multicolumn{1}{c|}{1 $\quad |\quad$ 1} & 4.5e${-5}$ $|$ 4.5e${-5}$ & \multicolumn{1}{c|}{1 $\quad |\quad$ 1} & 2.6e${-4}$ $|$ 2.6e${-4}$ & \multicolumn{1}{c|}{1 $\quad |\quad$ 1}     & 9.8e${-3}$ $|$ 9.8e${-3}$ \\ 
\hline
\end{tabular}}
\label{ttest2}

\label{proba}
\end{table}
%=======================================

\begin{table}[t]

\centering
% \caption{In the label-only scenario, the statistical significance comparison results of the forged watermark and the original watermark on CIFAR-10 (C10) and ImageNet (IM) by various backdoor watermarking methods. The red mark indicates that the forged watermark shows statistical significance.}
\caption{Statistical significance comparison (label-only API) on CIFAR-10 and ImageNet. Format: original $|$ forged.}
% \begin{subtable}
\resizebox{0.75\textwidth}{!}{
\centering
\begin{tabular}{c|c|c|cc|cc}
% \toprule[1.3pt]
\hline
\multirow{3}{*}{Data} & \multirow{3}{*}{M}   & BW    & \multicolumn{2}{c|}{Badnets}  & \multicolumn{2}{c}{Blended}   \\ 
\cline{3-7} 
      &  & Shape  & \multicolumn{1}{c|}{Cross} & Line  & \multicolumn{1}{c|}{Cross} & Line  \\ 
\cline{3-7} 
      &  & Metric & \multicolumn{1}{c|}{$p$-value} & $p$-value  & \multicolumn{1}{c|}{$p$-value} & $p$-value  \\ 
\hline
\multirow{4}{*}{C10} & \multirow{2}{*}{Res} & S   & \multicolumn{1}{c|}{0 $\quad |\quad$ 0} & $\quad \;$ 0 $\quad | $  1.5e${-2}$     & \multicolumn{1}{c|}{0 $\quad |\quad$ 0}          & 7.4e${-3}$ $|\quad$ \color{red}{0} $\quad \;$      \\ 
\cline{3-7} 
        &  & I & \multicolumn{1}{c|}{1 $\quad |\quad$ 1} & 1 $\quad |\quad$ 1  & \multicolumn{1}{c|}{1 $\quad |\quad$ 1}  & 1 $\quad |\quad$ 1  \\
\cline{2-7} 
        & \multirow{2}{*}{VGG} & S & \multicolumn{1}{c|}{0 $\quad |\quad$ 0} & 1.5e${-3}$ $|$ 3.0e${-3}$ & \multicolumn{1}{c|}{0 $\quad |\quad$ 0} & 5.5e${-3}$ $|\quad$ \color{red}{0} $\quad \;$  \\ 
\cline{3-7} 
        &  & I   & \multicolumn{1}{c|}{1 $\quad |\quad$ 1} & 1 $\quad |\quad$ 1 & \multicolumn{1}{c|}{1 $\quad |\quad$ 1} & 1 $\quad |\quad$ 1 \\ \hline
\multirow{4}{*}{IM}  & \multirow{2}{*}{Res} & S   & \multicolumn{1}{c|}{0 $\quad |\quad$ 0} & 5.5e${-3}$ $|$ 5.5e${-3}$ & \multicolumn{1}{c|}{9.9e${-3}$ $|$ \color{red}{7.4e${-3}$}} & 1 $\quad |\quad$ 1 \\ 
\cline{3-7} 
        &  & I   & \multicolumn{1}{c|}{1 $\quad |\quad$ 1} & 1 $\quad |\quad$ 1 & \multicolumn{1}{c|}{1 $\quad |\quad$ 1}  & 1 $\quad |\quad$ 1  \\
\cline{2-7} 
        & \multirow{2}{*}{VGG} & S   & \multicolumn{1}{c|}{0 $\quad |\quad$ 0} & 5.5e${-3}$ $|\quad$ \color{red}{0} $\quad \;$      & \multicolumn{1}{c|}{9.3e${-3}$ $|$ \color{red}{3.0e${-3}$}} & 9.9e${-3}$ $|$ \color{red}{3.0e${-3}$} \\
\cline{3-7} 
        &  & I   & \multicolumn{1}{c|}{1 $\quad |\quad$ 1} & 1 $\quad |\quad$ 1 & \multicolumn{1}{c|}{1 $\quad |\quad$ 1} & 1 $\quad |\quad$ 1 \\ 
\hline
% \bottomrule[1.0pt]
\end{tabular}}
\label{wtest1}
\vspace{1.0pt}

% \begin{subtable}

\resizebox{0.75\textwidth}{!}{
\centering
\begin{tabular}{c|c|c|c|c|c|c}
% \toprule[1.3pt]
\hline
\multirow{2}{*}{Data} & \multirow{2}{*}{M}   & BW    & $l_0$ inv               & Nature             & Trojan\_sq         & Trojan\_wm         \\ \cline{3-7} 
                      &                      & Metric & $p$-value                 & $p$-value            & $p$-value            & $p$-value            \\ \hline
\multirow{4}{*}{C10} & \multirow{2}{*}{Res} & S   & 0 $\quad |\quad$ 0      & 0 $\quad |\quad$ 0 & 0 $\quad |\quad$ 0 & 0 $\quad |\quad$ 0 \\ \cline{3-7} 
                      &                      & I   & 1 $\quad |\quad$ 1      & 1 $\quad |\quad$ 1 & 1 $\quad |\quad$ 1 & 1 $\quad |\quad$ 1 \\ \cline{2-7} 
                      & \multirow{2}{*}{VGG} & S   & $\quad \;$ 0 $\quad |$ 1.5e${-3}$ & 0 $\quad |\quad$ 0 & 0 $\quad |\quad$ 0 & 0 $\quad |\quad$ 0 \\ \cline{3-7} 
                      &                      & I   & 1 $\quad |\quad$ 1      & 1 $\quad |\quad$ 1 & 1 $\quad |\quad$ 1 & 1 $\quad |\quad$ 1 \\ \hline
\multirow{4}{*}{IM}  & \multirow{2}{*}{Res} & S   & 0 $\quad |\quad$ 0      & 0 $\quad |\quad$ 0 & 0 $\quad |\quad$ 0 & 0 $\quad |\quad$ 0 \\ \cline{3-7} 
                      &                      & I   & 1 $\quad |\quad$ 1      & 1 $\quad |\quad$ 1 & 1 $\quad |\quad$ 1 & 1 $\quad |\quad$ 1 \\ \cline{2-7} 
                      & \multirow{2}{*}{VGG} & S   & 0 $\quad |\quad$ 0      & 0 $\quad |\quad$ 0 & 0 $\quad |\quad$ 0 & 0 $\quad |\quad$ 0 \\ \cline{3-7} 
                      &                      & I   & 1 $\quad |\quad$ 1      & 1 $\quad |\quad$ 1 & 1 $\quad |\quad$ 1 & 1 $\quad |\quad$ 1 \\ 
\hline
% \bottomrule[1.0pt]
\end{tabular}}
\label{wtest2}
\label{label-only}
\end{table}

%=========================================
\begin{table*}[t]
% \caption{Comparison of the benign accuracy (BA ) (\%), the success rate of the original watermark (OWSR ) (\%), and the success rate of the forged watermark (FWSR ) (\%) of the backdoor watermarking method on the CIFAR-10 (C10) and ImageNet (IM) datasets, where Resnet-18 (Res) and VGG-19 (VGG) are two different suspicious models, and None represents that the model runs on the benign dataset without adding a watermark.} 
\caption{Classification performance: benign accuracy (BA), original watermark success rate (OWSR), and forged watermark success rate (FWSR) in \%.}
% \begin{subtable}

\centering
\resizebox{0.85\textwidth}{!}{
\begin{tabular}{c|c|c|cccccc|cccccc}
% \toprule[1.3pt]
\hline
 &
  BW &
   &
  \multicolumn{6}{c|}{Badnets} &
  \multicolumn{6}{c}{Blended} \\ \cline{2-2} \cline{4-15}
 &
  Shape &
  \multirow{-2}{*}{None} &
  \multicolumn{3}{c|}{Cross} &
  \multicolumn{3}{c|}{Line} &
  \multicolumn{3}{c|}{Cross} &
  \multicolumn{3}{c}{Line} \\ \cline{2-15} 
\multirow{-3}{*}{Data} &
  Metric &
  BA &
  \multicolumn{1}{c|}{BA} &
  \multicolumn{1}{c|}{OWSR} &
  \multicolumn{1}{c|}{FWSR} &
  \multicolumn{1}{c|}{BA} &
  \multicolumn{1}{c|}{OWSR} &
  FWSR &
  \multicolumn{1}{c|}{BA} &
  \multicolumn{1}{c|}{OWSR} &
  \multicolumn{1}{c|}{FWSR} &
  \multicolumn{1}{c|}{BA} &
  \multicolumn{1}{c|}{OWSR} &
  FWSR \\ \hline
 &
  Res &
  {91.4} &
  \multicolumn{1}{c|}{{91.9}} &
  \multicolumn{1}{c|}{{100}} &
  \multicolumn{1}{c|}{{100}} &
  \multicolumn{1}{c|}{91.7} &
  \multicolumn{1}{c|}{99.8} &
  99.9 &
  \multicolumn{1}{c|}{{91.6}} &
  \multicolumn{1}{c|}{{100}} &
  \multicolumn{1}{c|}{{100}} &
  \multicolumn{1}{c|}{91.3} &
  \multicolumn{1}{c|}{93.9} &
  99.0 \\ \cline{2-15} 
\multirow{-2}{*}{C10} &
  VGG &
  {91.7} &
  \multicolumn{1}{c|}{{91.6}} &
  \multicolumn{1}{c|}{{100}} &
  \multicolumn{1}{c|}{{100}} &
  \multicolumn{1}{c|}{91.6} &
  \multicolumn{1}{c|}{99.6} &
  98.4 &
  \multicolumn{1}{c|}{{91.2}} &
  \multicolumn{1}{c|}{{99.8}} &
  \multicolumn{1}{c|}{{100}} &
  \multicolumn{1}{c|}{90.2} &
  \multicolumn{1}{c|}{93.1} &
  97.0 \\ \hline
 &
  Res &
  {87.0} &
  \multicolumn{1}{c|}{{86.7}} &
  \multicolumn{1}{c|}{{99.8}} &
  \multicolumn{1}{c|}{{99.8}} &
  \multicolumn{1}{c|}{86.0} &
  \multicolumn{1}{c|}{96.4} &
  95.6 &
  \multicolumn{1}{c|}{{86.3}} &
  \multicolumn{1}{c|}{{90.9}} &
  \multicolumn{1}{c|}{{95.3}} &
  \multicolumn{1}{c|}{85.4} &
  \multicolumn{1}{c|}{81.0} &
  86.9 \\ \cline{2-15} 
\multirow{-2}{*}{IM} &
  VGG &
  {90.6} &
  \multicolumn{1}{c|}{{89.4}} &
  \multicolumn{1}{c|}{{100}} &
  \multicolumn{1}{c|}{{99.9}} &
  \multicolumn{1}{c|}{89.8} &
  \multicolumn{1}{c|}{97.7} &
  98.4 &
  \multicolumn{1}{c|}{{89.5}} &
  \multicolumn{1}{c|}{{95.6}} &
  \multicolumn{1}{c|}{{98.8}} &
  \multicolumn{1}{c|}{89.7} &
  \multicolumn{1}{c|}{89.3} &
  94.4 \\ 
  \hline
% \bottomrule[1.0pt]
\end{tabular}}
%\caption{The comparison results of Badnets and Blended,  with Cross and Line denoting different style patterns of original watermarks.}
\label{ba1}

% \end{subtable}

\vspace{1.0pt}

% \begin{subtable}

\centering
\resizebox{0.85\textwidth}{!}{
\begin{tabular}{c|c|c|ccc|ccc|ccc|ccc}
% \toprule[1.3pt]
\hline
 &
  BW &
  None &
  \multicolumn{3}{c|}{$l_0$ inv} &
  \multicolumn{3}{c|}{Nature} &
  \multicolumn{3}{c|}{Trojan\_sq} &
  \multicolumn{3}{c}{Trojan\_wm} \\ \cline{2-15} 
\multirow{-2}{*}{Data} &
  Metric &
  BA &
  \multicolumn{1}{c|}{BA} &
  \multicolumn{1}{c|}{OWSR} &
  FWSR &
  \multicolumn{1}{c|}{BA} &
  \multicolumn{1}{c|}{OWSR} &
  FWSR &
  \multicolumn{1}{c|}{BA} &
  \multicolumn{1}{c|}{OWSR} &
  FWSR &
  \multicolumn{1}{c|}{BA} &
  \multicolumn{1}{c|}{OWSR} &
  FWSR \\ \hline
 &
  Res &
  {93.5} &
  \multicolumn{1}{c|}{{92.9}} &
  \multicolumn{1}{c|}{{100}} &
  {100} &
  \multicolumn{1}{c|}{{93.5}} &
  \multicolumn{1}{c|}{{100}} &
  {100} &
  \multicolumn{1}{c|}{{92.6}} &
  \multicolumn{1}{c|}{{99.8}} &
  {100} &
  \multicolumn{1}{c|}{{93.2}} &
  \multicolumn{1}{c|}{{100}} &
  {100} \\ \cline{2-15} 
\multirow{-2}{*}{C10} &
  VGG &
  {93.0} &
  \multicolumn{1}{c|}{{92.9}} &
  \multicolumn{1}{c|}{{100}} &
  {99.2} &
  \multicolumn{1}{c|}{{91.7}} &
  \multicolumn{1}{c|}{{100}} &
  {98.9} &
  \multicolumn{1}{c|}{{92.2}} &
  \multicolumn{1}{c|}{{99.9}} &
  {100} &
  \multicolumn{1}{c|}{{92.8}} &
  \multicolumn{1}{c|}{{100}} &
  {100} \\ \hline
 &
  Res &
  {88.2} &
  \multicolumn{1}{c|}{{89.8}} &
  \multicolumn{1}{c|}{{100}} &
  {100} &
  \multicolumn{1}{c|}{{89.7}} &
  \multicolumn{1}{c|}{{99.8}} &
  {99.4} &
  \multicolumn{1}{c|}{{89.8}} &
  \multicolumn{1}{c|}{{100}} &
  {99.6} &
  \multicolumn{1}{c|}{{89.5}} &
  \multicolumn{1}{c|}{{100}} &
  {99.7} \\ \cline{2-15} 
\multirow{-2}{*}{IM} &
  VGG &
  {91.2} &
  \multicolumn{1}{c|}{{92.7}} &
  \multicolumn{1}{c|}{{100}} &
  {100} &
  \multicolumn{1}{c|}{{92.8}} &
  \multicolumn{1}{c|}{{100}} &
  {100} &
  \multicolumn{1}{c|}{{93.4}} &
  \multicolumn{1}{c|}{{100}} &
  {100} &
  \multicolumn{1}{c|}{{93.1}} &
  \multicolumn{1}{c|}{{100}} &
  {100} \\ 
\hline
% \bottomrule[1.0pt]
\end{tabular}}
\label{ba2}
\label{water_ba}
\end{table*}

\begin{figure}[t]
    \centering
    \includegraphics[width=0.85\textwidth]{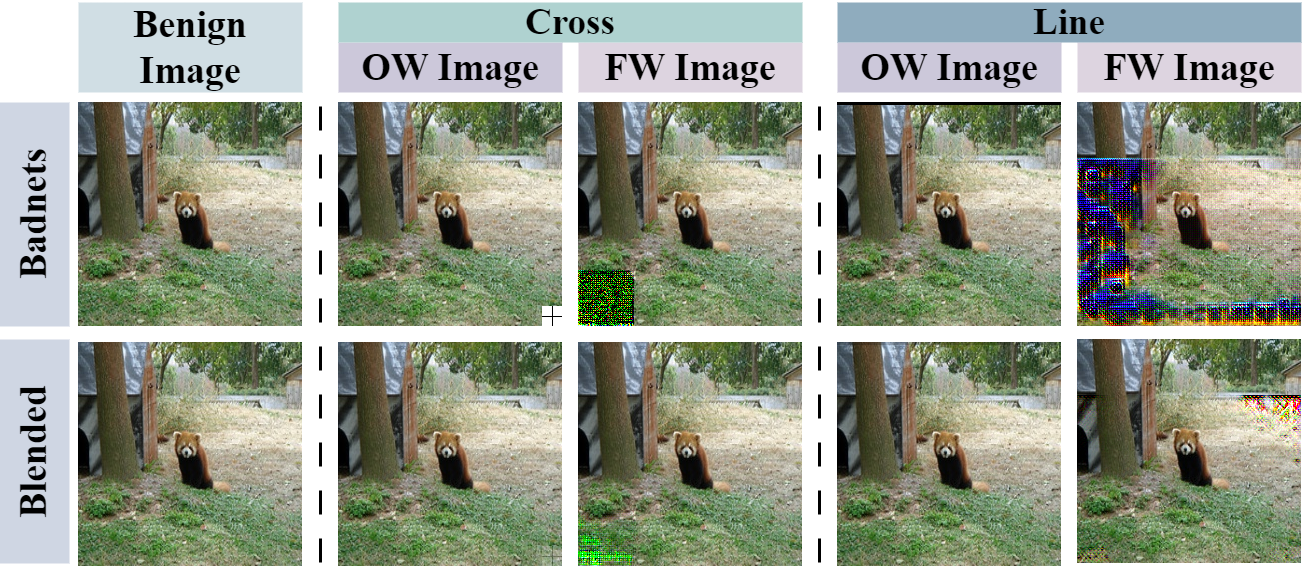}
    % \caption{Examples of benign, original watermarked (OW), and forged watermarked (FW) images using the Badnets and Blended backdoor watermark attack methods, where the crosses and lines represent different styles of the original watermark, respectively.
    % }
    \caption{Visual comparison of benign, original watermarked (OW), and forged watermarked (FW) images for BadNets and Blended methods.}
    \label{fig:watermark-example}
\end{figure}

\begin{figure*}[htb]
\centering
    
    % \begin{subfigure}
    %     \includegraphics[width=0.32\textwidth]{MSE_Comparison2.png}
    % \end{subfigure}
    % \hfill
    % \begin{subfigure}
    %     \includegraphics[width=0.32\textwidth]{PSNR_Comparison2.png}
    % \end{subfigure}
    % \hfill
    % \begin{subfigure}
    %     \includegraphics[width=0.32\textwidth]{SSIM_Comparison2.png}
    % \end{subfigure}

    \subfloat{%
        \includegraphics[width=0.3\textwidth]{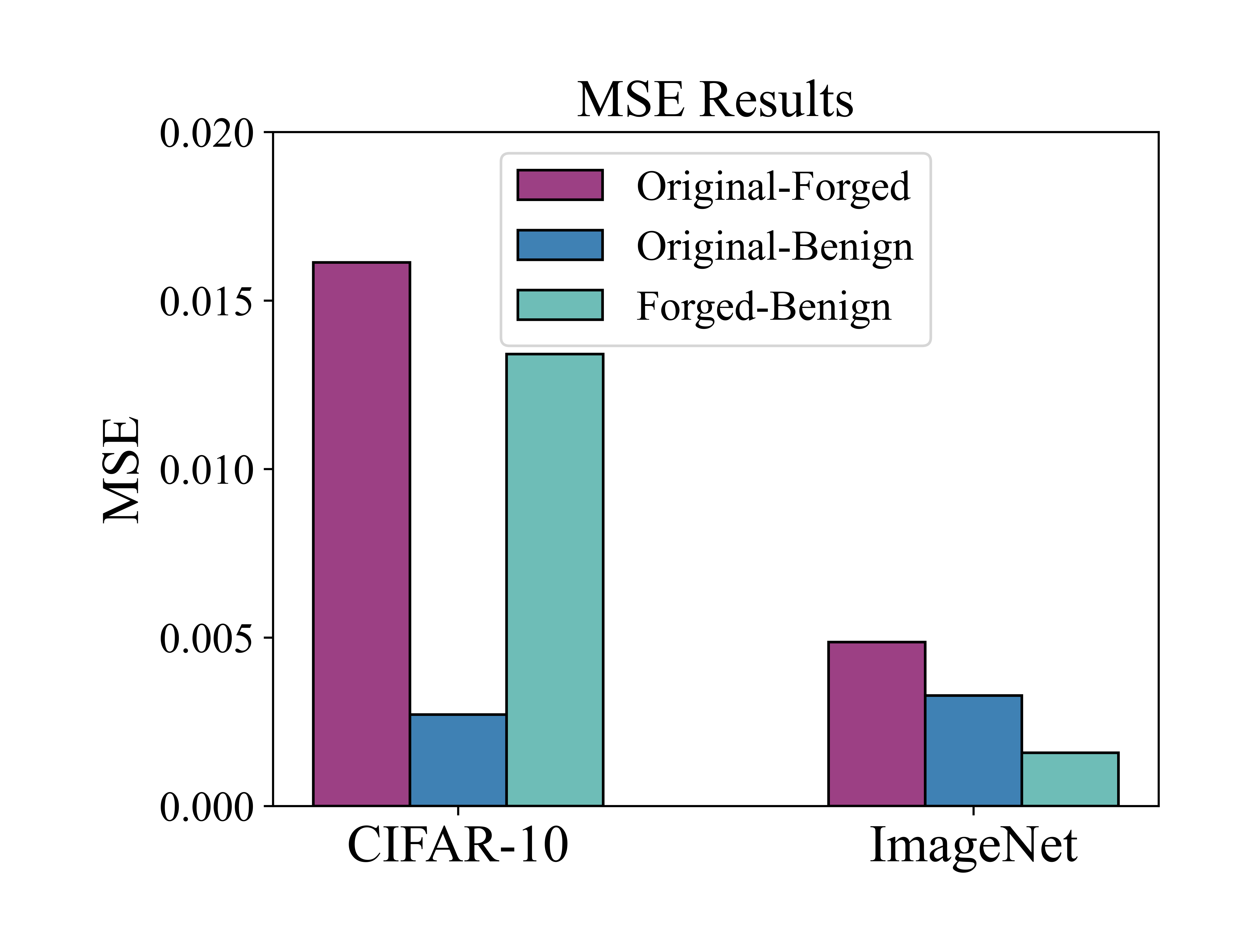}%
        \label{fig:mse}
    }
    \hfill
    \subfloat{%
        \includegraphics[width=0.3\textwidth]{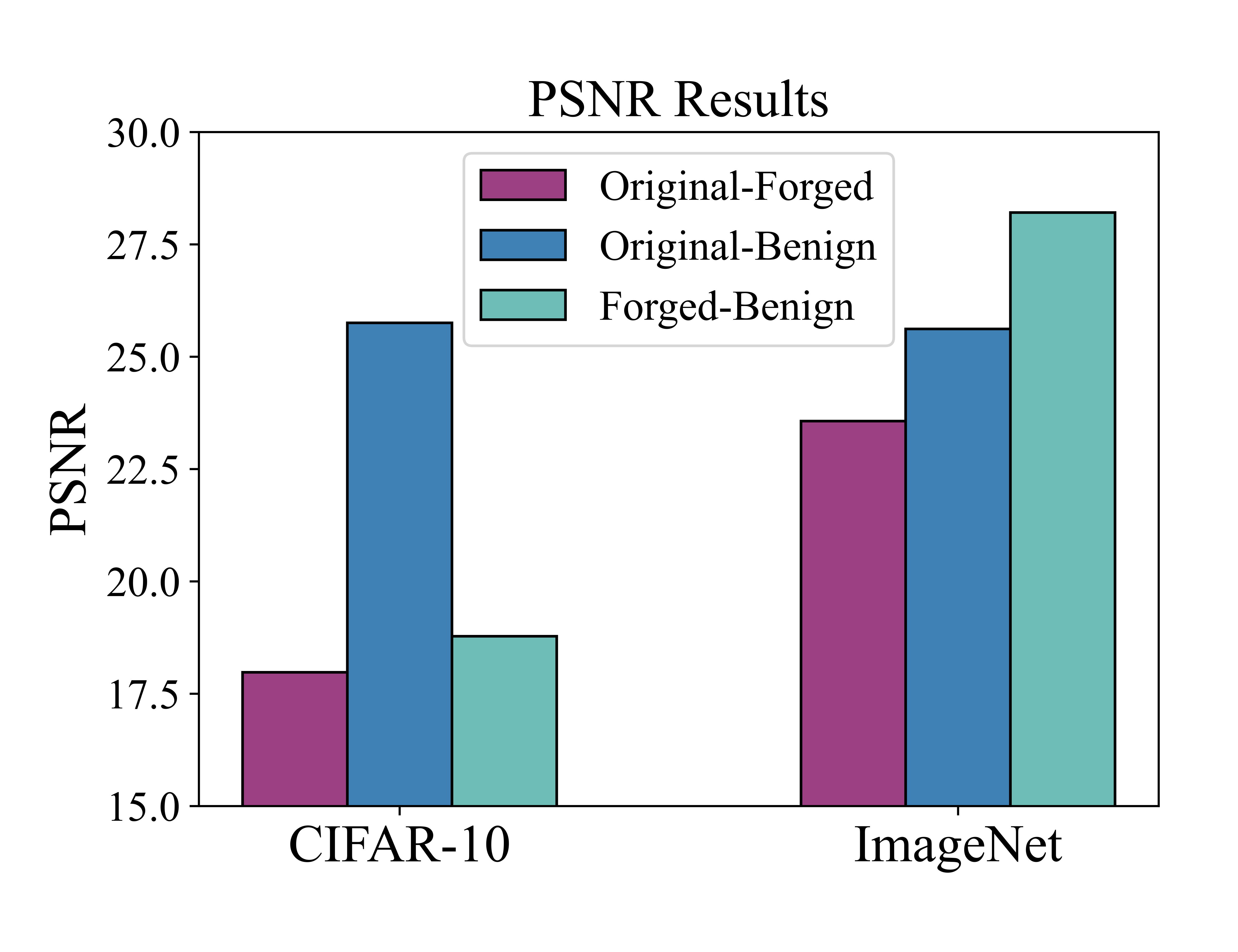}%
        \label{fig:psnr}
    }
    \hfill
    \subfloat{%
        \includegraphics[width=0.3\textwidth]{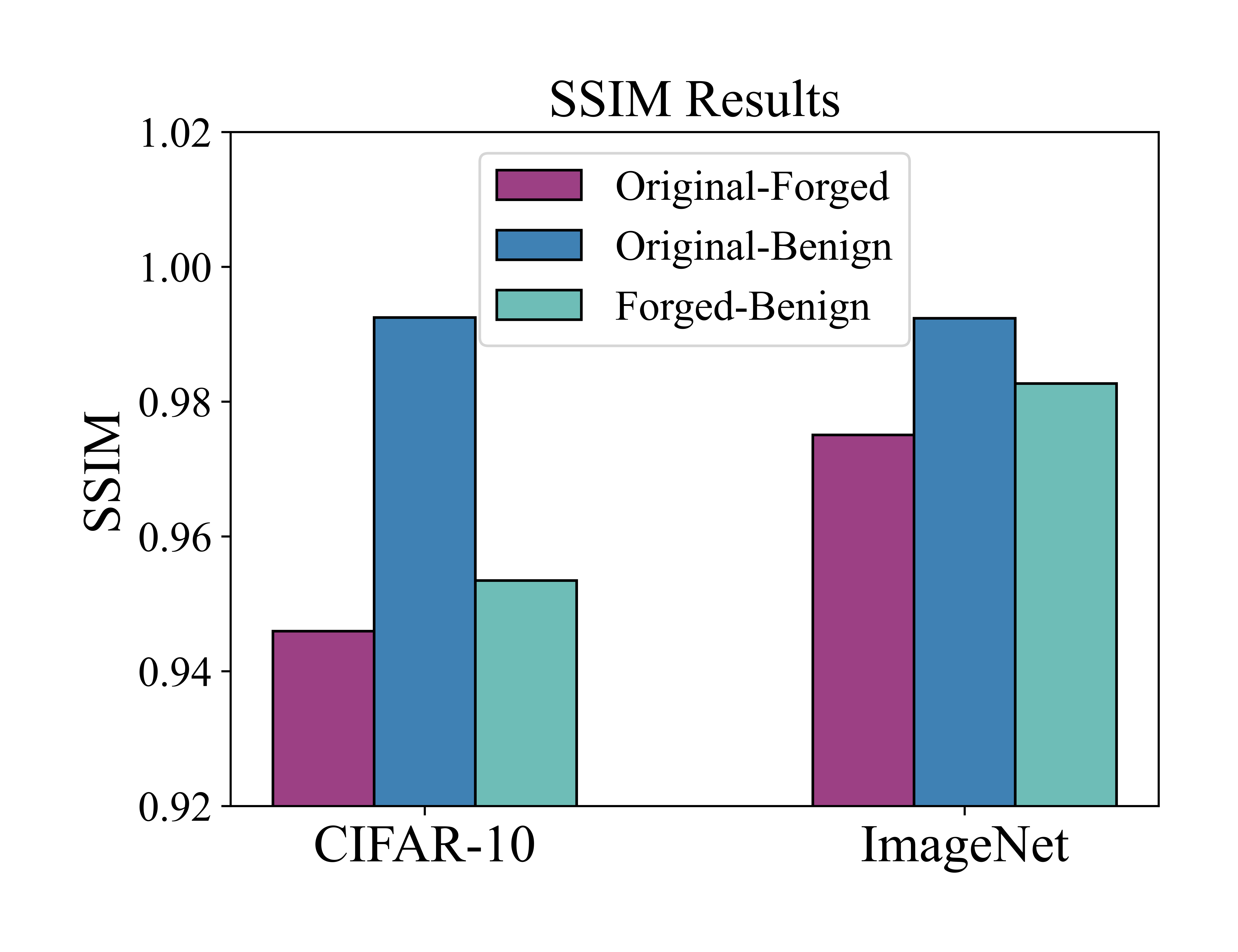}%
        \label{fig:ssim}
    }
     % 第一行的标题
% \caption{Comparisons of PSNR, SSIM, and MSE metrics using the Badnets method and ResNet-18 models  trained on the CIFAR-10 and ImageNet datasets as target models, with Original-Forged, Original-Benign, and Forged-Benign representing the comparative pairs of original watermarked images and forged watermarked images, original watermarked images and benign images, and forged watermarked images and benign images, respectively.}
\caption{Image quality metrics (PSNR, SSIM, MSE) between original and forged watermarks on CIFAR-10 and ImageNet using BadNets with ResNet-18.}
\label{fig:image-quality}
\end{figure*}

\subsection{Statistical Equivalence Analysis (RQ2)}

To demonstrate the consistency of forged watermarks and original watermarks in copyright verification in \textbf{RQ2}, we adopt various backdoor watermarking methods in two types of APIs (probability available and label-only) and two scenarios (independent model and stealing model) to verify the equivalence between the original watermark and the forged watermark, as shown in Fig. \ref{fig:watermark-example}. For BadNet and Blended methods, we follow the protocol in \cite{li2023backdoorbox} to embed \textbf{Cross} (the small block in the lower right corner) and \textbf{Line} (the thick black line at the top) into the image to obtain the watermarked image. For other backdoor watermarking methods, we conduct experiments according to the description in the original literature \cite{li2020invisible,chen2017targeted,liu2018trojaning}.

\noindent \textbf{Hypothesis Testing Results.} For the case where the classification model outputs probability (i.e., probability available API), we perform T-tests on the original watermark dataset and the forged watermark dataset respectively, and determine the dataset infringement based on the significance level of the statistical results (Table \ref{proba}). Similarly, when the classification model only outputs labels (label-only API), the Wilcoxon signed rank test is performed on the original watermark dataset and the forged watermark dataset to obtain the p-values, as shown in Table \ref{label-only}.

% In the stealing model scenario, we use a one-sided T-test. When $\Delta P$ is larger and the $p$-value is smaller, the probability of rejecting the null hypothesis is greater. Unlike the stealing model, for the independent model scenario, when we use a one-sided T-test, the smaller the $\Delta P$ value and the larger the corresponding $p$-value, the greater the probability of accepting the null hypothesis. 

According to the theory of hypothesis testing, we can draw the following conclusions from Table \ref{proba} and \ref{label-only}

\im{

\item The $p$-value in the stealing model or the $1-p$ value in the independent model is much smaller than the significance level of $0.05$ usually set in statistics.

\item In the probability available scenario, both the original watermark and the forged watermark can accurately determine with high confidence that the dataset has been stolen. Here, the $p$-value in the theft model scenario is close to 0, $\Delta P$ $\gg $ 0, while in the independent model scenario, the $p$-value $\gg $ 0.05, $\Delta P$ is close to 0.

\item In the probability available scenario, whether in the stealing model scenario or the independent model scenario, the $p$-value and $\Delta P$ of the forged watermark are the same as or even higher than the original watermark, as shown in the red marks in Table \ref{proba}. This shows that in copyright verification, the forged watermark shows at least the same or even stronger statistical significance as the original watermark.

}

\begin{table*}[htb]
% \caption{In the probability available scenario, the impact of loss functions $\mc{L}_{W}$, $\mc{L}_{B}$ and $\mc{L}_{BW}$ on different forgery watermarking methods on the datasets CIFAR-10 (C10) and ImageNet (IM), where we also consider different suspicious models Res and VGG.} 
\caption{Ablation study on loss components (probability-available API). Format: $\mathcal{L}_W | \mathcal{L}_B | \mathcal{L}_{BW}$.}
% \begin{subtable}{\textwidth}
\centering
\resizebox{0.85\textwidth}{!}{%
\begin{tabular}{c|c|c|cccc}
% \toprule[1.3pt]
\hline
\multirow{3}{*}{Data} & \multirow{3}{*}{M}   & BW    & \multicolumn{4}{c}{Badnets} \\ 
\cline{3-7} 
    &  & Shape & \multicolumn{2}{c|}{Cross} & \multicolumn{2}{c}{Line} \\ 
    \cline{3-7} 
    &  & Metric & \multicolumn{1}{c|}{$p$-value ($\mc{L}_{W} | \mc{L}_{B} | \mc{L}_{WB}$) } & \multicolumn{1}{c|}{$\Delta P$ ($\mc{L}_{W} | \mc{L}_{B} | \mc{L}_{WB}$)} & \multicolumn{1}{c|}{$p$-value ($\mc{L}_{W} | \mc{L}_{B} | \mc{L}_{WB}$)} & $\Delta P$ ($\mc{L}_{W} | \mc{L}_{B} | \mc{L}_{WB}$)\\ 
\hline
\multirow{4}{*}{C10} & \multirow{2}{*}{Res} & S   & \multicolumn{1}{c|}{2.3e${-172}$ $|\quad$ 1 $\quad|$ 1.1e${-173}$} & \multicolumn{1}{c|}{9.9e${-1}$ $\, | \;$ 2.9e${-4}$ $\; | \,$ 9.9e${-1}$}    & \multicolumn{1}{c|}{$\ $ 1.6e${-69}$ $|\quad$ 1 $\quad \,|$ 1.8e${-101}$}  & 8.8e${-1}$ $|$ -3.4e${-4}$ $|$ 9.8e${-1}$   \\ 
\cline{3-7} 
    &  & I   & \multicolumn{1}{c|}{1 $\quad | \quad$ 1 $\quad |\quad$ 1}  & \multicolumn{1}{c|}{7.9e${-4}$ $\, | \;$ 7.9e${-4}$ $\; | \,$ 7.9e${-4}$}    & \multicolumn{1}{c|}{1 $\quad | \quad$ 1 $\quad |\quad$ 1} & -3.6e${-4}$ $|$ -3.6e${-4}$ $|$ -3.6e${-4}$ \\ 
\cline{2-7} 
    & \multirow{2}{*}{VGG} & S   & \multicolumn{1}{c|}{7.8e${-140}$ $|\quad$ 1 $\quad|$ 1.1e${-143}$} & \multicolumn{1}{c|}{1.7e${-1}$ $|$ -2.7e${-4}$ $\, |$ 9.9e${-1}$}   & \multicolumn{1}{c|}{8.6e${-205}$ $|\, \quad$ 1 $\quad|$ 3.2e${-75}$ $\ $}  & 9.9e${-1}$ $|$ -3.1e${-4}$ $|$ 9.7e${-1}$   \\ 
\cline{3-7} 
    &  & I   & \multicolumn{1}{c|}{1 $\quad | \quad$ 1 $\quad |\quad$ 1}  & \multicolumn{1}{c|}{5.4e${-5}$ $\, | \;$ 5.4e${-5}$ $\; | \,$ 5.4e${-5}$}    & \multicolumn{1}{c|}{1 $\quad | \quad$ 1 $\quad |\quad$ 1} & -5.8e${-5}$ $|$ -5.8e${-5}$ $|$ -5.8e${-5}$ \\ 
\hline
\multirow{4}{*}{IM}  & \multirow{2}{*}{Res} & S   & \multicolumn{1}{c|}{9.0e${-169}$ $|\quad$ 1 $\quad|$ 1.4e${-177}$} & \multicolumn{1}{c|}{9.9e${-1}$ $|$ -1.2e${-5}$ $\, | $ 9.9e${-1}$}   & \multicolumn{1}{c|}{1.8e${-118}$ $|\, \quad$ 1 $\quad| $ 4.3e${-63}$ $\ $}  & 9.8e${-1}$ $|$ -2.2e${-4}$ $|$ 9.3e${-1}$   \\
\cline{3-7} 
    &  & I   & \multicolumn{1}{c|}{1 $\quad | \quad$ 1 $\quad |\quad$ 1}  & \multicolumn{1}{c|}{8.2e${-6}$ $\, | \;$ 8.2e${-6}$ $\; | \,$ 8.2e${-6}$}    & \multicolumn{1}{c|}{1 $\quad | \quad$ 1 $\quad |\quad$ 1}  & 5.2e${-6}$ $\, | \;$ 5.2e${-6}$ $\; | \,$ 5.2e${-6}$    \\ 
\cline{2-7} 
    & \multirow{2}{*}{VGG} & S   & \multicolumn{1}{c|}{1.8e${-181}$ $|\quad$ 1 $\quad|$ 2.2e${-181}$} & \multicolumn{1}{c|}{9.9e${-1}$ $|$ -3.3e${-4}$ $\, |$ 9.9e${-1}$}   & \multicolumn{1}{c|}{2.3e${-300}$ $|\quad$ 1 $\quad|$ 3.8e${-112}$} & 9.9e${-1}$ $|$ -2.1e${-6}$ $|$ 9.7e${-1}$   \\ 
\cline{3-7} 
    &  & I   & \multicolumn{1}{c|}{1 $\quad | \quad$ 1 $\quad |\quad$ 1}  & \multicolumn{1}{c|}{-1.9e${-7}$ $|$ -1.9e${-7}$ $|$ -1.9e${-7}$} & \multicolumn{1}{c|}{1 $\quad | \quad$ 1 $\quad |\quad$ 1}   & 2.7e${-6}$ $\, | \;$ 2.7e${-6}$ $\; | \,$ 2.7e${-6}$    \\ 
\hline
% \bottomrule[1.0pt]
\end{tabular}
}
%\caption{Comparison results of different loss functions on Badnets.} 
\label{abla-ttest-badnets}
% \end{subtable}

\vspace{1.0pt}

% \begin{subtable}{\textwidth}
\centering
\resizebox{0.85\textwidth}{!}{%
\begin{tabular}{c|c|c|cccc}
% \toprule[1.3pt]
\hline
\multirow{3}{*}{Data} & \multirow{3}{*}{M}   & BW & \multicolumn{4}{c}{Blended}  \\ 
\cline{3-7} 
    &  & Shape & \multicolumn{2}{c|}{Cross} & \multicolumn{2}{c}{Line} \\ 
\cline{3-7} 
    &   & Metric & \multicolumn{1}{c|}{$p$-value ($\mc{L}_{W} | \mc{L}_{B} | \mc{L}_{WB}$)} & \multicolumn{1}{c|}{$\Delta P$ ($\mc{L}_{W} | \mc{L}_{B} | \mc{L}_{WB}$)} & \multicolumn{1}{c|}{$p$-value 
 ($\mc{L}_{W} | \mc{L}_{B} | \mc{L}_{WB}$)} & $\Delta P$ ($\mc{L}_{W} | \mc{L}_{B} | \mc{L}_{WB}$) \\ 
\hline
\multirow{4}{*}{C10} & \multirow{2}{*}{Res} & S   & \multicolumn{1}{c|}{1.3e${-164}$ $|\quad$ 1 $\quad|$ 9.5e${-165}$} & \multicolumn{1}{c|}{9.9e${-1}$ $\, | \;$  2.5e${-3}$ $\; | \,$ 9.9e${-1}$}  & \multicolumn{1}{c|}{1.0e${-161}$ $|\quad$ 1 $\quad|$ 1.4e${-121}$} & 9.9e${-1}$ $\, | \;$ 4.7e${-3}$ $\; | \,$ 9.8e${-1}$    \\ 
\cline{3-7} 
     &   & I   & \multicolumn{1}{c|}{1 $\quad | \quad$ 1 $\quad |\quad$ 1}            & \multicolumn{1}{c|}{6.2e${-5}$ $\, | \;$ 6.2e${-5}$ $\; | \,$ 6.2e${-5}$}  & \multicolumn{1}{c|}{1 $\quad | \quad$ 1 $\quad |\quad$ 1}   & -2.3e${-4}$ $|$ -2.3e${-4}$ $|$ -2.3e${-4}$ \\ 
\cline{2-7} 
    & \multirow{2}{*}{VGG} & S   & \multicolumn{1}{c|}{2.7e${-241}$ $|\quad$ 1 $\quad|$ 3.7e${-241}$} & \multicolumn{1}{c|}{9.9e${-1}$ $\, | \;$ 1.2e${-4}$ $\; | \,$ 9.9e${-1}$}  & \multicolumn{1}{c|}{3.9e${-68}$ $|\quad$ 1 $\quad|$ 1.5e${-96}$}   & 4.1e${-1}$ $\, | \;$ 6.3e${-3}$ $\; | \,$ 9.7e${-1}$    \\ 
\cline{3-7} 
     &   & I   & \multicolumn{1}{c|}{1 $\quad | \quad$ 1 $\quad |\quad$ 1}            & \multicolumn{1}{c|}{1.8e${-4}$ $\, | \;$ 1.8e${-4}$ $\; | \,$ 1.8e${-4}$}  & \multicolumn{1}{c|}{1 $\quad | \quad$ 1 $\quad |\quad$ 1}   & -1.9e${-4}$ $|$ -1.9e${-4}$ $|$ -1.9e${-4}$ \\ 
\hline
\multirow{4}{*}{IM}  & \multirow{2}{*}{Res} & S   & \multicolumn{1}{c|}{$\,$ 1.2e${-52}$ $\; |\quad$ 1 $ \quad|$ 1.35e${-65}$}  & \multicolumn{1}{c|}{8.7e${-1}$ $\, | \;$ 2.3e${-5}$ $\; | \,$ 8.8e${-1}$}  & \multicolumn{1}{c|}{8.1e${-48}$ $|\quad$ 1 $\quad|$ 2.4e${-25}$}   & 7.8e${-1}$ $|$ -1.9e${-4}$ $|$ 6.5e${-1}$   \\ 
\cline{3-7} 
    &  & I   & \multicolumn{1}{c|}{1 $\quad | \quad$ 1 $\quad |\quad$ 1}  & \multicolumn{1}{c|}{5.0e${-6}$ $\, | \;$ 5.0e${-6}$ $\; | \,$ 5.0e${-6}$}  & \multicolumn{1}{c|}{1 $\quad | \quad$ 1 $\quad |\quad$ 1}            & -2.4e${-7}$ $|$ -2.4e${-7}$ $|$ -2.2e${-7}$ \\ 
\cline{2-7} 
    & \multirow{2}{*}{VGG} & S   & \multicolumn{1}{c|}{2.3e${-125}$ $|\quad$ 1 $\quad| \,$ 2.1e${-77}$ $\,$}  & \multicolumn{1}{c|}{9.7e${-1}$ $|$ -1.1e${-4}$ $|$ 9.5e${-1}$} & \multicolumn{1}{c|}{1.7e${-177}$ $|\quad\,$ 1 $\quad|\;\;$ 7.6e${-72}$}  & 9.8e${-1}$ $|$ -9.6e${-6}$ $|$ 9.5e${-1}$   \\ 
\cline{3-7} 
    &  & I   & \multicolumn{1}{c|}{1 $\quad | \quad$ 1 $\quad |\quad$ 1}            & \multicolumn{1}{c|}{6.2e${-8}$ $\, | \;$ 6.2e${-8}$ $\; | \,$ 6.7e${-8}$}  & \multicolumn{1}{c|}{1 $\quad | \quad$ 1 $\quad |\quad$ 1}            & 2.7e${-6}$ $\, | \;$ 2.7e${-6}$ $\; | \,$ 2.7e${-6}$    \\ 
\hline
% \bottomrule[1.0pt]
\end{tabular}}
%\caption{Comparison results of different loss functions on Blended.} 
% \end{subtable}
\label{abla-ttest}
\end{table*}

\begin{table*}[ht]
% \caption{In the label-only scenario, ablation study of loss functions $\mc{L}_{W}$, $\mc{L}_{B}$ and $\mc{L}_{BW}$ on CIFAR-10 (C10) and ImageNet (IM).}
\caption{Ablation study on loss components (label-only API). Format: $\mathcal{L}_W | \mathcal{L}_B | \mathcal{L}_{BW}$.}
\centering
\resizebox{0.85\textwidth}{!}{

\begin{tabular}{c|c|c|cc|cc}
% \toprule[1.3pt]
\hline
\multirow{3}{*}{Data} & \multirow{3}{*}{M}   & BW    & \multicolumn{2}{c|}{Badnets}   & \multicolumn{2}{c}{Blended}  \\ 
\cline{3-7} 
    &  & Shape    & \multicolumn{1}{c|}{Cross}  & Line  & \multicolumn{1}{c|}{Cross}  & Line   \\ 
\cline{3-7} 
    &  & Metric & \multicolumn{1}{c|}{$p$-value ($\mc{L}_{W} | \mc{L}_{B} | \mc{L}_{WB}$)}  & $p$-value ($\mc{L}_{W} | \mc{L}_{B} | \mc{L}_{WB}$)  & \multicolumn{1}{c|}{$p$-value ($\mc{L}_{W} | \mc{L}_{B} | \mc{L}_{WB}$)}  & $p$-value ($\mc{L}_{W} | \mc{L}_{B} | \mc{L}_{WB}$)  \\ 
\hline
\multirow{4}{*}{C10} & \multirow{2}{*}{Res} & S   & \multicolumn{1}{c|}{0 $\quad | \quad$ 1 $\quad |\quad$ 0}      & 4.3e${-2}$ $| \quad$ 1 $\quad |$ 1.5e${-1}$ & \multicolumn{1}{c|}{0 $\quad | \quad$ 1 $\quad |\quad$ 0}          & 0 $\quad | \quad$ 1 $\quad | \quad$ 0  \\ 
\cline{3-7} 
    &  & I   & \multicolumn{1}{c|}{1 $\quad | \quad$ 1 $\quad |\quad$ 1}  & 1 $\quad | \quad$ 1 $\quad |\quad$ 1   & \multicolumn{1}{c|}{1 $\quad | \quad$ 1 $\quad |\quad$ 1}  & 1 $\quad | \quad$ 1 $\quad |\quad$ 1 \\ 
\cline{2-7} 
    & \multirow{2}{*}{VGG} & S   & \multicolumn{1}{c|}{0 $\quad | \quad$ 1 $\quad |\quad$ 0} & 3.0e${-3}$ $\quad | \quad$ 1 $\quad | \quad$ 3.0e${-3}$     & \multicolumn{1}{c|}{0 $\quad | \quad$ 1 $\quad |\quad$ 0}          & 5.5e${-3}$ $\quad | \quad$ 1 $\quad | \quad$ 0 $\qquad$ \\
\cline{3-7} 
    &  & I   & \multicolumn{1}{c|}{1 $\quad | \quad$ 1 $\quad |\quad$ 1}      & 1 $\quad | \quad$ 1 $\quad |\quad$ 1          & \multicolumn{1}{c|}{1 $\quad | \quad$ 1 $\quad |\quad$ 1}          & 1 $\quad | \quad$ 1 $\quad |\quad$ 1       \\ 
\hline
\multirow{4}{*}{IM}  & \multirow{2}{*}{Res} & S   & \multicolumn{1}{c|}{0 $\quad | \quad$ 1 $\quad |\quad$ 0}      & 5.5e${-3}$ $\quad | \quad$ 1 $\quad |\quad$ 5.5e${-3}$          & \multicolumn{1}{c|}{9.3e${-3}$ $| \quad$ 1 $\quad |$ 7.4e${-3}$} & 7.4e${-1}$ $\quad | \quad$ 1 $\quad | \quad $ 1 $\qquad$ \\ 
\cline{3-7} 
    &  & I   & \multicolumn{1}{c|}{1 $\quad | \quad$ 1 $\quad |\quad$ 1}      & 1 $\quad | \quad$ 1 $\quad |\quad$ 1          & \multicolumn{1}{c|}{1 $\quad | \quad$ 1 $\quad |\quad$ 1}          & 1 $\quad | \quad$ 1 $\quad |\quad$ 1       \\ 
\cline{2-7} 
    & \multirow{2}{*}{VGG} & S   & \multicolumn{1}{c|}{0 $\quad | \quad$ 1 $\quad |\quad$ 0}      & 0 $\quad | \quad$ 1 $\quad |\quad$ 0          & \multicolumn{1}{c|}{7.1e${-3}$ $\quad | \quad$ 1 $\quad |\quad$ 3.0e${-3}$}          & 9.9e${-3}$ $\quad | \quad$ 1 $\quad | \quad$ 3.0e${-3}$  \\ 
\cline{3-7} 
    &  & I   & \multicolumn{1}{c|}{1 $\quad | \quad$ 1 $\quad |\quad$ 1}      & 1 $\quad | \quad$ 1 $\quad |\quad$ 1          & \multicolumn{1}{c|}{1 $\quad | \quad$ 1 $\quad |\quad$ 1}          & 1 $\quad | \quad$ 1 $\quad |\quad$ 1       \\ 
\hline
% \bottomrule[1.0pt]
\end{tabular}}

\label{abla-wtest}
\end{table*}

%========================================abla-FWSR

\begin{table*}[htb]
% \caption{Ablation study of the loss function ($\mc{L}_{B},\mc{L}_{W},\mc{L}_{BW}$) on the Forged Watermark Success Rate (FWSR) on different datasets including CIFAR-10 (C10) and ImageNet (IM) and different backdoor watermarking methods (Badnets and Blended).}
\caption{Ablation study: impact of loss components on FWSR (\%). Format: $\mathcal{L}_W | \mathcal{L}_B | \mathcal{L}_{BW}$.}
\centering
\resizebox{0.85\textwidth}{!}{
\begin{tabular}{c|c|cc|cc}
% \toprule[1.3pt]
\hline
\multirow{3}{*}{Data} & BW   & \multicolumn{2}{c|}{Badnets} & \multicolumn{2}{c}{Blended} \\ 
\cline{2-6} 
    & Shape   & \multicolumn{1}{c|}{Cross} & Line               & \multicolumn{1}{c|}{Cross} & Line \\ 
\cline{2-6} 
    & Metric & \multicolumn{1}{c|}{FWSR ($\mc{L}_{W} | \mc{L}_{B} | \mc{L}_{WB}$)} & FWSR ($\mc{L}_{W} | \mc{L}_{B} | \mc{L}_{WB}$) & \multicolumn{1}{c|}{FWSR ($\mc{L}_{W} | \mc{L}_{B} | \mc{L}_{WB}$)} & FWSR ($\mc{L}_{W} | \mc{L}_{B} | \mc{L}_{WB}$)  \\ 
\hline
\multirow{2}{*}{C10} & Res  & \multicolumn{1}{c|}{100 $\quad | \quad$ 1.4 $\quad |\quad$ 100}   & 98.3 $\quad | \quad$ 0.9 $\quad |\quad$ 99.9 $\,$ & \multicolumn{1}{c|}{100 $\quad | \quad$ 0.9 $\quad |\quad$ 100}   & 100 $\quad | \quad$ 1.1 $\quad |\quad$ 99.0  \\ 
\cline{2-6} 
    & VGG  & \multicolumn{1}{c|}{$\ $ \textcolor{red}{16.5} $\quad | \,\quad$ 1.3 $\quad |\quad$ \textcolor{blue}{100} $\;\,$}  & 100 $\quad | \quad$ 1.2 $\quad |\quad$ 98.4  & \multicolumn{1}{c|}{100 $\quad | \quad$ 1.3 $\quad |\quad$ 100}   & \textcolor{red}{45.3} $\quad | \quad$ 1.3 $\quad |\quad$ \textcolor{blue}{97.0} $\,$\\ 
\hline
\multirow{2}{*}{IM}  & Res  & \multicolumn{1}{c|}{$\;$ 99.5 $\quad | \quad$ 0.2 $\quad |\quad$ 99.8 $\,$} & 100 $\quad | \quad$ 0.4 $\quad |\quad$ 95.6  & \multicolumn{1}{c|}{96.1 $\quad | \quad$ 0.2 $\quad |\quad$ 95.3} & 95.5 $\quad | \quad$ 2.0 $\quad |\quad$ 86.9 $\,$\\ 
\cline{2-6} 
    & VGG  & \multicolumn{1}{c|}{$\,$ 100 $\quad | \quad$ 0.1 $\quad |\quad$ 99.9}  & 100 $\quad | \quad$ 0.1 $\quad |\quad$ 98.4  & \multicolumn{1}{c|}{99.5 $\quad | \quad$ 0.1 $\quad |\quad$ 98.8} & 100 $\quad | \quad$ 0.1 $\quad |\quad$ 94.4  \\
\hline
% \bottomrule[1.0pt]
\end{tabular}}
\label{abla_fwsr}
\end{table*}

\noindent \textbf{Classification Performance.} We analyze the BA (benign accuracy) and WSR (watermark success rate) of the two watermarks. From the results in Table \ref{water_ba}, we can see that compared with the performance of the watermarked model on the benign image test set, the performance on the watermarked image test set usually decreases by less than 1\%, and in some cases it is slightly improved, which further proves that watermarking technology does not hinder the use of normal datasets. It is worth noting that the success rate of the forged watermark (FWSR) is very close to or exceeds the success rate of the original watermark (OWSR). For example, on the Blended Line watermark dataset, the success rate of the forged watermark is a significant improvement over the original watermark (86.9\% vs. 81.0\%). These results highlight the effectiveness of the forged watermark in simulating the original watermark function.

\begin{figure}[htb!]
    \centering
    \includegraphics[width=0.60\textwidth]{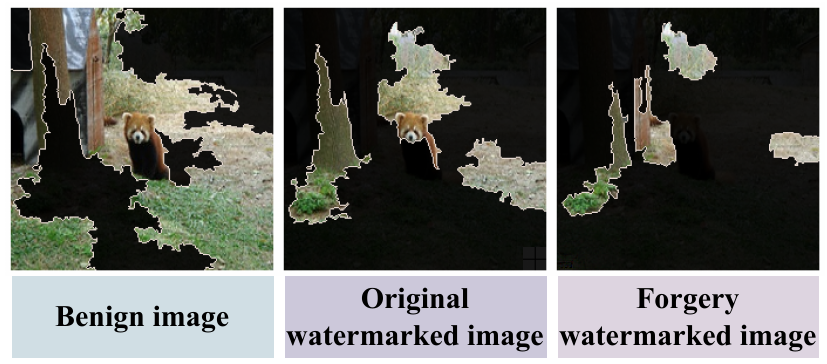}
    % \caption{LIME provides diagnostic interpretation of images with and without different watermarks, where regions contributing to the watermark classification are explicitly marked.}
    \caption{LIME interpretability analysis showing distinct attention regions for benign, original, and forged watermarked images.}
    \label{fig:lime}
\end{figure}

\noindent \textbf{Visual Distinctiveness.} To quantitatively evaluate the visual difference between the original and forged watermarks, we used three main image quality metrics: PSNR, SSIM, and MSE. Generally, lower PSNR and SSIM values indicate larger differences, while larger MSE indicates larger differences between the pixels of the two images. We quantitatively analyzed the differences between the benign, original, and forged watermarked images, as shown in Fig. \ref{fig:image-quality}, which shows that there are significant differences between the original and forged watermarked images. In addition, the visualization results shown in Fig. \ref{fig:watermark-example} also clearly demonstrate the obvious differences in style between the original and forged watermarks.

Furthermore, we employ Locally Interpretable Model-Agnostic Explanations (LIME) \cite{ribeiro2016should} to explain the differences between the original and forged watermarks and evaluate the effectiveness of the attack. Fig. \ref{fig:lime} illustrates how LIME identifies regions that affect the watermarked model's predictions. When analyzing benign images, the model focuses on the object itself. However, for watermarked images, the focus clearly shifts to the peripheral regions. Notably, the focus regions associated with the original watermark are significantly different from those associated with the forged watermark.

\subsection{Ablation Study on Losses}

To answer the impact of the loss function submitted in \textbf{RQ3} on FW-Gen, we conducted ablation experiments from two perspectives: hypothesis testing and classification.

\noindent \textbf{Hypothesis Testing.} We conduct ablation experiments on the loss function of {FW-Gen} using Badnets and Blended methods to evaluate the impact of two different loss pairs on the hypothesis testing results, as shown in Tables \ref{abla-ttest} and \ref{abla-wtest}. We find that \tf{for hypothesis testing, $\mc{L}_\tm{W}$ loss is more important than $\mc{L}_\tm{B}$}, where the $p$-value of $\mc{L}_\tm{B}$ is always equal to $1$, which is consistent with our intuition.
Since the watermarked images (including original watermarks and forged watermarks) and benign images perform similarly on the benign model, the loss $\mc{L}_\tm{B}$ has little impact on the benign model.

\noindent \textbf{Classification.} We further analyze the impact of the two loss functions on the classification performance, as shown in Table \ref{abla_fwsr}. We find that although the distillation loss $\mc{L}_{\tm{W}}$ on the watermarked model plays an important role in most cases, there are two exceptions, as shown in the red and blue fonts in Table \ref{abla_fwsr}. On the VGG-19 model, the success rate of watermark attack using only loss $\mc{L}_{\tm{W}}$ is only $16.5\%$ and $45.3\%$, while after adding loss $\mc{L}_{\tm{B}}$, the success rate of watermark attack is greatly improved, reaching $100\%$ and $97\%$ respectively. This indicates that \textbf{$\mc{L}_{\tm{B}}$ also plays an important role in improving WSR compared with $\mc{L}_{\tm{W}}$}.

%==============================================================================
% DEFENSE DISCUSSION
%==============================================================================
\section{Defense Discussion and Limitations}
\label{sec:defense}

Our findings expose fundamental vulnerabilities in current backdoor watermarking schemes. We discuss potential countermeasures and acknowledge limitations.

\subsection{Potential Defenses}

\noindent \textbf{Cryptographic Timestamping.} The most direct defense establishes temporal precedence through cryptographic commitment. Dataset owners can register a hash of their watermark pattern on a blockchain~\cite{waheed2024grove} or with a trusted timestamping authority before release. This creates an immutable record proving the owner's watermark existed before any potential forgery.

\noindent \textbf{Steganographic Watermarking.} More sophisticated embedding techniques using imperceptible perturbations~\cite{li2020invisible,doan2021backdoor} may evade detection. However, as Table~\ref{tab:detection} shows, even advanced methods achieve high detection accuracy.

\noindent \textbf{Multi-Watermark Schemes.} Dataset owners could embed multiple independent watermarks with different patterns and target labels, exponentially increasing attack complexity.

\noindent \textbf{Behavioral Diversity.} Watermarks could induce complex behavioral signatures (\eg, specific confidence distributions, multi-class activation patterns) that are harder to replicate without the original design.

\subsection{Limitations}
\noindent \textbf{Detection Dependency:} If watermark detection accuracy falls below $\sim$80\%, forgery quality degrades.

\noindent \textbf{Model Access Requirement:} The attacker requires query access to their deployed model.

\noindent \textbf{Computational Overhead:} FW-Gen training requires approximately 500 iterations with both models.

\noindent \textbf{Single-Watermark Assumption:} Our attack targets single watermarks; extending to multi-watermark schemes requires additional research.

%==============================================================================
% CONCLUSION
%==============================================================================
\section{Conclusion}
\label{sec:conclusion}

This paper challenges the prevailing assumption that backdoor watermarking provides reliable dataset ownership verification. By adopting an adversarial perspective, we demonstrate that accused attackers can forge watermarks that are statistically indistinguishable from original watermarks in DOV procedures, effectively producing counter-evidence to refute infringement claims. Our proposed FW-Gen framework achieves this through a novel distillation-based approach that transfers the behavioral characteristics of original watermarks to forged ones while maintaining visual distinctiveness. Comprehensive experiments consistently demonstrate that forged and original watermarks exhibit equivalent statistical significance, undermining the evidentiary value of DOV results. These findings carry important implications: (1) DOV results alone are insufficient for copyright disputes---additional mechanisms such as cryptographic timestamps are necessary; (2) Our analysis motivates the development of forgery-resistant watermarking schemes. We hope this work stimulates further research into robust dataset protection mechanisms capable of withstanding adversarial scrutiny.

\newpage
\bibliographystyle{splncs04}
\bibliography{mybib}
\appendix
\section{Training Flow}
Algorithm~\ref{alg:fwgen} summarizes the complete FW-Gen training procedure.

\begin{algorithm}[t]
\caption{FW-Gen Training}
\label{alg:fwgen}
\begin{algorithmic}[1]
\REQUIRE Protected dataset $\Dp$, detected watermark $\tow$, target label $\yhat$, benign model $f$, suspicious model $\ftilde$
\ENSURE Trained FW-Gen parameters $(\phi, \varrho)$
\STATE Extract clean dataset $D_c$ from $\Dp$ by removing watermarked samples
\STATE Initialize encoder $\mathcal{E}_\phi$ and decoder $\mathcal{D}_\varrho$
\FOR{iteration $= 1$ to max\_iterations}
    \STATE Sample batch $\{x_i\}_{i=1}^{B}$ from $D_c$
    \STATE Sample $\epsilon_1, \epsilon_2 \sim \mathcal{N}(0, I)$
    \STATE $\mu, \sigma \leftarrow \mathcal{E}_\phi(\epsilon_1)$
    \STATE $z \leftarrow \mu + \sigma \odot \epsilon_2$
    \STATE $\tfw \leftarrow \mathcal{D}_\varrho(z)$
    \STATE Construct $x_{\text{ow}} = \mathcal{G}(x_i, \tow)$, $x_{\text{fw}} = \mathcal{G}(x_i, \tfw)$
    \STATE Compute $\LB$ using Eq.~(6) with model $f$
    \STATE Compute $\LW$ using Eq.~(7) with model $\ftilde$
    \STATE $\mathcal{L} \leftarrow \LB + \LW$
    \STATE Update $\phi, \varrho$ via gradient descent on $\mathcal{L}$
\ENDFOR
\RETURN $\phi, \varrho$
\end{algorithmic}
\end{algorithm}

\end{document}